\documentclass[12pt,preprint]{aastex}

\usepackage{multicol, multirow}
\usepackage{amsmath,amssymb,graphics,cite,hyperref,twoopt,lineno,color,soul,lscape,chngpage,booktabs}
\usepackage[caption=false]{subfig}

\begin{document}

\title{A search for spectral hysteresis and energy-dependent time lags from X-ray and TeV gamma-ray observations of Mrk 421}

\author{
A.~U.~Abeysekara\altaffilmark{1},
S.~Archambault\altaffilmark{2},
A.~Archer\altaffilmark{3},
W.~Benbow\altaffilmark{4},
R.~Bird\altaffilmark{5},
M.~Buchovecky\altaffilmark{5},
J.~H.~Buckley\altaffilmark{3},
V.~Bugaev\altaffilmark{3},
J.~V~Cardenzana\altaffilmark{6},
M.~Cerruti\altaffilmark{4},
X.~Chen\altaffilmark{7,8}
% ...
}
\altaffiltext{1}{Department of Physics and Astronomy, University of Utah, Salt Lake City, UT 84112, USA}
\altaffiltext{2}{Physics Department, McGill University, Montreal, QC H3A 2T8, Canada}
\altaffiltext{3}{Department of Physics, Washington University, St. Louis, MO 63130, USA}
\altaffiltext{4}{Fred Lawrence Whipple Observatory, Harvard-Smithsonian Center for Astrophysics, Amado, AZ 85645, USA}
\altaffiltext{5}{Department of Physics and Astronomy, University of California, Los Angeles, CA 90095, USA}
\altaffiltext{6}{Department of Physics and Astronomy, Iowa State University, Ames, IA 50011, USA}
\altaffiltext{7}{Institute of Physics and Astronomy, University of Potsdam, 14476 Potsdam-Golm, Germany}
\altaffiltext{8}{DESY, Platanenallee 6, 15738 Zeuthen, Germany}
% ...

\begin{abstract}
Blazars are variable emitters across all wavelengths over a wide range of timescales, from months down to minutes. 
It is therefore essential to observe blazars simultaneously at different wavelengths, especially in the X-ray and gamma-ray bands, where the broadband spectral energy distributions usually peak. 

In this work, we report on three ``target-of-opportunity'' (ToO) observations of Mrk 421, one of the brightest TeV blazars, triggered by a strong flaring event at TeV energies in 2014. These observations feature long, continuous, and simultaneous exposures with {\it XMM-Newton} (covering X-ray and optical/ultraviolet bands) and VERITAS (covering TeV gamma-ray band), along with contemporaneous observations from other gamma-ray facilities (MAGIC and {\it Fermi}-LAT) and a number of radio and optical facilities. 
Although neither rapid flares nor significant X-ray/TeV correlation are detected, these observations reveal subtle changes in the X-ray spectrum of the source over the course of a few days. 
We search the simultaneous X-ray and TeV data for spectral hysteresis patterns and time delays, which could provide insight into the emission mechanisms and the source properties (e.g. the radius of the emitting region, the strength of the magnetic field, and related timescales). The observed broadband spectra are consistent with a one-zone synchrotron self-Compton model. 
We find that the power spectral density distribution at $\gtrsim 4\times 10^{-4}$~Hz from the X-ray data can be described by a power-law model with an index value between 1.2 and 1.8, and do not find evidence for a steepening of the power spectral index (often associated with a characteristic length scale) compared to the previously reported values at lower frequencies. 

\end{abstract}

\section{Introduction}
Relativistic outflows in the form of bipolar jets are an important means of carrying energy away from many accreting compact objects in astrophysics. Such objects range from X-ray binaries of a few solar masses, to the bright central regions with black holes of millions of solar masses in some galaxies, known as active galactic nuclei (AGN). Blazars, an extreme sub-class of the AGN family, are oriented such that one of the relativistic jets is pointed almost directly at the observer, resulting in a bright, point-like source \citep[e.g.][]{Padovani95}. 

The spectral energy distribution (SED) of blazars typically exhibits two peaks in the $\nu F_{\nu}$ representation \citep[e.g.][]{Fossati98}. The lower-energy peak in the SED of blazars is commonly associated with synchrotron radiation from relativistic electrons/positrons (electrons hereafter) in the jet. The higher-energy peak could be the result of inverse-Compton scattering from the same electrons \citep[in leptonic models, e.g.][]{Marscher1985, Maraschi1992, Bottcher1998}, or of radiation from hadronic processes, e.g. $\pi^0$ decay \citep[e.g.][]{Sahu2013}, photopion processes \citep[e.g.][]{Mannheim1991, Dimitrakoudis2014}, or proton synchrotron emission \citep[e.g.][]{Aharonian2000}. 
The current instruments are usually unable to measure the broadband SED with the necessary energy coverage and time resolution \citep[e.g.][]{Bottcher13}, therefore variability plays a crucial role in distinguishing between these models \citep[e.g.][]{Mastichiadis13}. 

Blazars are variable emitters across all wavelengths over a wide range
of timescales. On long timescales (days to months), radio
observations with high angular resolution have suggested a connection
between knots with distinct polarization angles and outbursts of radio
flux, sometimes with an optical and/or gamma-ray counterpart
\citep[e.g.][]{Rani15}. Correlated multiwavelength (MWL) variability
studies are important for investigating the particles and magnetic
field in the jets, as well as their spatial structure
\citep[e.g.,][]{Blazejowski05,Katarzynski05,Arlen13}. For example, in
synchrotron self-Compton (SSC) models for high-frequency-peaked BL Lac objects (HBLs), X-ray and very-high-energy (VHE; 100~GeV - 100~TeV) fluxes
are highly correlated and most strongly variable when the electron
injection rate changes. A general correlation between the X-ray and TeV
fluxes on longer timescales has been observed with no systematic
lags. However, \citet{Fossati08} found ``an intriguing hint'' that the
correlation between X-ray and TeV fluxes may be different for
variability with different timescales. Specifically, the data suggest
a roughly quadratic dependence of the VHE flux on the X-ray flux for
timescales of hours, but a less steep, close to linear relationship,
for timescales of days \citep[e.g.][]{Fossati08, Aleksic14M42010, Balokovic16}.

On shorter timescales, blazar variability has been observed in both X-ray and gamma-ray bands \citep[e.g.,][]{Gaidos96, Cui04, Pryal15}. Especially interesting are the fast TeV flares with doubling times as short as a few minutes, the production mechanisms of which are even less well understood than the variability on longer timescales. 
One major obstacle to understanding such flares lies in the practical challenge in organizing simultaneous MWL observations on short timescales. 
Firstly, it is difficult, if not impossible, to predict when a blazar will flare, due to the stochastic nature of its emission. Secondly, it takes time to coordinate target-of-opportunity (ToO) observations with X-ray satellites and ground-based telescopes in response to a spontaneous flaring event. Thirdly, most of the current X-ray satellites have relatively short orbital periods, and are frequently interrupted by Earth occultation and the South Atlantic Anomaly (SAA) passage, while observations from ground-based Cherenkov telescopes may be affected by the weather, or precluded by daylight. These gaps in the observations increase the chances of missing a fast flare and introduce bias into timing analyses (e.g. cross correlation and power spectrum). 
The {\it XMM-Newton} satellite has a long orbital period (48 hr), capable of providing observations of $>10$~hr with no exposure gaps. It is therefore uniquely well-suited for monitoring and studying sub-hour variability, and is chosen as the primary X-ray instrument in this work. 
It is also worth noting that fast automated analyses of multi-wavelength data from TeV gamma-ray blazars are done regularly, providing the potential to deploy ToO observations at short notice if a strong flare is detected from a blazar.

Spectral hysteresis and energy-dependent time lags observed in blazars have also provided unique insights into the different timescales associated with particle acceleration and energy loss \citep[e.g.][]{KRM98,Bottcher02}, which can then be used to test different blazar models. 
However, such studies have been limited to X-ray observations, as a large number of photons are needed to provide a constraining result \citep[e.g.][]{Takahashi96, Kataoka00, Cui04, Falcone04}. The increased sensitivity of the current generation of Cherenkov telescopes, such us VERITAS and MAGIC, has motivated the search of fast TeV gamma-ray variability and hysteresis of blazars in this work. 

Within the framework of a 6-month long multi-instrument campaign, the MAGIC telescopes observed on 2014 April 25 a VHE gamma-ray flux reaching eight times the flux above 300~GeV of the Crab Nebula (Crab units, C.~U.) from the TeV blazar Mrk 421 \citep[e.g.][]{Punch92}, which is about 16 times brighter than usual. 
This triggered a joint ToO program by {\it XMM-Newton}, VERITAS, and MAGIC.
Three, approximately 4-hour long, continuous and simultaneous observations in both X-ray and TeV gamma-ray bands were carried out on Apr 29, May 1, and May 3, 2014. 
This was the third time in eight years that Mrk~421 triggered the joint ToO program. Compared to the last two triggers in 2006 and 2008 \citep{Acciari11M4flare08}, the source flux observed by VERITAS was significantly higher at 1 - 2.5 C.U. in 2014. 
In this work, we focus on the simultaneous VERITAS-{\it XMM-Newton} data obtained from the ToO observations in 2014 (listed in Table~\ref{MWLobs}), and complement this study with other contemporaneous MWL observations (including those of MAGIC) of Mrk~421. The details of the large flare observed with MAGIC on 2014 April 25 will be reported elsewhere. 

\begin{table}
    \centering
    \caption{Summary of the simultaneous ToO observations of Mrk 421 in 2014. 
    Column 1 and 2 are the UTC and MJD dates of the observations, respectively. Column 3 and 4 are the start and end time of the VERITAS and {\it XMM-Newton} observations, respectively. }
    { \small
    \begin{tabular}{cccc} \\ \hline\hline
      UTC Date & MJD & VERITAS   & {\it XMM}-EPN    \\ \hline 
      2014-04-29 & 56776 & 03:19-08:02 &04:24-08:00   \\ 
      2014-05-01 & 56778 & 03:24-06:10 &03:46-07:53   \\  
      2014-05-03 & 56780 & 03:31-06:05 &03:35-07:42   \\ \hline 
      \end{tabular} \\
      \label{MWLobs}
    }    
\end{table}

\section{Observations and Data Analysis}
\label{sec:obs_data}
\subsection{VERITAS and MAGIC}
\label{subsec:VHE}
VERITAS is an array of four 12-m ground-based imaging atmospheric Cherenkov telescopes in southern Arizona, each equipped with a camera consisting of 499 photomultiplier tubes (PMTs) \citep{Holder11}. It is sensitive to gamma rays in the energy range from $\sim$100~GeV to $\sim$30~TeV with an energy resolution of $\sim$15\%, and covers a 3.5$^{\circ}$ field-of-view with an angular resolution (68\% containment) of $\sim$0.1$^{\circ}$. It is capable of making a detection at a statistical significance of 5 standard deviations (5 $\sigma$) of a point source of 1\% C.U. in $\sim$25~hours. 
The systematic uncertainty on the energy calibration is estimated at 20\%, and that on the spectral index is estimated at 0.2 \citep{Madhavan13}. 

VERITAS has been monitoring Mrk~421 regularly for approximately 20 hours every year, as part of several long-term MWL monitoring campaigns \citep[e.g.][]{Acciari11, Aleksic14M42010}. The general strategy is to take a 30-minute exposure on every third night when the source is visible, with coordinated, simultaneous X-ray observations (usually with the X-Ray Telescope (XRT) on board the {\it Swift} satellite). 
In contrast, the three long and simultaneous observations with {\it XMM-Newton} and VERITAS on 2014 Apr 29, May 1, and May 3 are specific attempts to catch rapid flares on top of elevated flux states simultaneously in X-ray and TeV bands. 

Due to high atmospheric dust conditions at the VERITAS site on May 1, only data from the ToO observations on Apr 29 and May 3 have been used. The VERITAS observations on these two nights were taken in ``wobble'' mode \citep{Fomin1994} with the source offset 0.5$^{\circ}$ from the center of the field-of-view. 
The zenith angles of the observations were between 10$^{\circ}$ and 40$^{\circ}$. 
After deadtime correction, the total exposure time from these observations is 6.14 hours. 
The data have been analyzed using the data analysis procedures described in \citet{Cogan08}.
Standard gamma-ray selection cuts, previously optimized for sources with a power-law spectrum of a photon index 2.5, have been applied to reject cosmic-ray (CR) background events. 
The reflected-region background model \citep{Berge07} was used to estimate the number of CR background events that have passed the cuts, and a generalized method from \citet{LiMa83} has been used for the calculation of statistical significance. The VERITAS results are shown in Table~\ref{Vobs}. 

To parameterize the curvature in the VERITAS-measured TeV spectra, a power-law model with an exponential cutoff has been used to fit the daily spectra: 
\begin{equation}
\label{PLexpModel}
\frac{dN}{dE} = K \left( \frac{E}{E_0} \right)^{-\alpha} e^{-\frac{E}{E_\text{cutoff}}}. 
\end{equation}
However, we used a power-law model in the hysteresis study in Section~\ref{XhystSec}, as it adequately describes each 10-min integrated spectrum without the cutoff energy as an extra degree of freedom. 

\begin{table}
  \centering
    \caption{Summary of VERITAS observations of Mrk 421 (the analysis details are given in Section~\ref{sec:res})}
    { 
    \begin{tabular}{cccccccc} \\ \hline\hline
      Date & Exposure & Significance & Non & Noff & $\alpha$ & Gamma-ray rate & Background rate \\ \hline 
      		     &   (minutes)   &  $\sigma$  & &  & & $\text{photons}\;\text{min}^{-1}$ & $\text{CRs}\;\text{min}^{-1}$ \\ \hline 
      2014-04-29 & 237.4 & 97.4 & 2481 & 538 & 0.1 & $ 10.2 \pm 0.2 $ & 0.21  \\
      2014-05-01 & 146.4 & -  & - & - & - & - & - \\ 
      2014-05-03 & 131.0 & 74.3 & 1443 & 315 & 0.1 & $ 10.8 \pm 0.3 $ & 0.22 \\ \hline 
      \end{tabular} 
            \label{Vobs}
    }
\end{table}

The Major Atmospheric Gamma-ray Imaging Cherenkov (MAGIC) telescope system consists of two 17-meter telescopes, located at the Observatory Roque de los Muchachos, on the Canary island of La Palma (28.8~N, 17.8~W, 2200~m a.s.l.). Stereoscopic observations provide a sensitivity of detecting a point source at $\sim$0.7\% C. U. above 220~GeV in 50 hours of observation, and allow measurement of photons in the energy range from 50~GeV to above 50~TeV. 
The night-to-night systematic uncertainty in the VHE flux measurement by MAGIC is estimated to be of the order of 11\% \citep{aleksic2016}. 

Mrk~421 was observed by MAGIC for six nights from 2014 April 28 to 2014 May 4, as part of a longer multiwavelength
observational campaign. The source was observed in ``wobble'' mode, with $0.4^\circ$
offset with respect to the nominal source position
\citep{Fomin1994}. After discarding data observed in poor weather
conditions, the total analyzed data amount to 3.3 hours of
observations, with exposures per observation ranging from 14 to 38 minutes, and zenith angles spanning from 9$^{\circ}$ to 42$^{\circ}$. 

The MAGIC data have been analyzed using the standard MAGIC analysis and reconstruction software~\citep{zanin2013}. 
The integral flux was computed above 560~GeV, the same as the energy threshold found in the VERITAS long-term light curve, in order to use all the observations including those at large zenith angles. The source gamma-ray flux varied between 1.3 and 2.2 C.U. above 560~GeV for different days in this period, with no significant intra-night variability. 
This flux value is 3-5 times larger than the typical VHE flux of
Mrk~421 \citep{acciari2014,aleksic2015}. These observations are not simultaneous with the {\it XMM-Newton} observations. The source is known to
change spectral index with flux level \citep{Krennrich02}, and hence we computed the photon flux above 560 GeV using the measured spectral shape above 400 GeV, which ranged from 2.8 to 3.3. 

\subsection {{\it Fermi}-LAT}
The {\it Fermi} Large Area Telescope (LAT) is a pair-conversion high-energy gamma-ray telescope covering an energy range from about 20 MeV to more than 300 GeV \citep{Atwood09}. It has a large field-of-view of 2.4~sr that covers the full sky every 3~hr in the nominal survey mode. Thus, {\it Fermi}-LAT provides long-term sampling of the entire sky. However, it has a small effective area of $\sim8000 \; \text{cm}^2$ for $>1$~GeV, which is usually not sufficient to resolve variability on timescales of hours or less. 

We analyzed the {\it Fermi}-LAT Pass 8 data in the week of the VERITAS observations and produced daily-averaged spectra and a daily-binned light curve. We selected events of class {\em source} and type {\em front+back} with an energy between 0.1 and 300\,GeV in a $10^{\circ}$ region of interest (RoI) centered at the location of Mrk~421, and removed events with a zenith angle $> 90^{\circ}$. The data were processed using the publicly available {\it Fermi}-LAT science tools (\texttt{v10r0p5}) with instrument response functions (\texttt{P8R2\_SOURCE\_V6}). A model with the contributions of all sources within the RoI with a test statistic value greater than 3, a list of 3FGL sources within a source region of a radius of $20^{\circ}$ from Mrk~421, and the contribution of the Galactic (using file \texttt{gll\_iem\_v06.fit}) and isotropic (using file \texttt{iso\_P8R2\_SOURCE\_V6\_v06.txt}) diffuse emission was used. This model has been fitted to LAT Pass 8 data between 2014 Apr 1 and 2014 June 1 using an unbinned likelihood analysis (\texttt{gtlike}). The test-statistic maps were examined to ensure no unmodeled transient sources were present in the region of interest during the period analyzed. All other best fit parameters in the model were then fixed, except the spectral normalization and the power-law index of Mrk~421, in order to perform a spectral and temporal maximum likelihood analysis. 

\subsection {{\it XMM-Newton} and {\it Swift}-XRT}
\label{sec:XMM}
The {\it XMM-Newton} satellite carries the European Photon Imaging Camera (EPIC) pn X-ray CCD camera \citep{Struder01}, including two Metal-Oxide-Silicon (MOS) cameras and a pn camera. The reflection grating spectrometers (RGS) with high energy resolution are installed in front of the MOS detector. The incoming X-ray flux is divided into two portions for the MOS and RGS detectors. The EPIC-pn (EPN) detector receives the unobstructed beam and is capable of observing with very high time resolution. The Optical/ultraviolet (UV) Monitor (OM) onboard the {\it XMM-Newton} satellite provides the capability to cover a $17'\times 17'$ square region between 170 nm and 650 nm \citep{Mason01}. The OM is equipped with six broad-band filters ({\it U}, {\it B}, {\it V}, {\it UVW1}, {\it UVM2} and {\it UVW2}).  

Three ToO observations were taken simultaneously with the VERITAS observations on Apr 29, May~1, and May~3, 2014. 
To fully utilize the high time resolution capability of {\it XMM-Newton} in both the X-ray and optical/UV bands, all three ToO observations of Mrk 421 were taken in EPN timing mode and OM fast mode. MOS and RGS were also operated during the observations, but the data have not been used due to the relatively low timing resolution and the lack of X-ray spectral lines from the source. The EPN camera covers a spectral range of approximately 0.5 - 10 keV and, with the {\it UVM2} filter, the OM covers the range of about 200 - 270 nm. 

{\it XMM-Newton} EPN and OM data have been analyzed using Statistical Analysis System (SAS) software version 13.5 \citep{Gabriel04}. 
An X-ray loading correction and a rate-dependent pulse height amplitude (RDPHA) correction were performed using the SAS tool \texttt{epchain}. 
We ran the SAS task \texttt{epproc} to produce the RDPHA results, which applies calibrations using known spectral lines and is likely more accurate than the alternative charge transfer inefficiency (CTI) corrections \footnote{See \url{http://xmm2.esac.esa.int/docs/documents/CAL-SRN-0312-1-4.pdf}}. 
Note that even after the RDPHA corrections, residual absorption features (not associated with the source) can still be present in the spectrum \citep[see e.g.][]{Pintore14}. 
To account for the source and the residual spectral features, the X-ray spectra were fitted using \texttt{XSPEC} version \texttt{12.8.1} with a model including a power law, a photoelectric absorption component representing the Galactic neutral hydrogen absorption, an absorption edge component, and two Gaussian components. The latter three components are only associated with the instrument. They account for the oxygen K line at $\sim$0.54~keV, the silicon K line at $\sim$1.84~keV, and the gold M line at $\sim$2.2~keV, respectively. 
The model can be expressed as follows: 
\begin{equation}
\label{XMMmodel}
\frac{dN}{dE} = 
\begin{cases}
e^{-n_H\sigma(E)} \left[K_{PL} E^{-\alpha}+ \sum_i \frac{K_{G,i}}{\sqrt{2\pi}\sigma_i}e^{-\frac{(E-E_{0,i})^2}{2\sigma_i^2}} \right],& E\leqslant E_{c};  \\
e^{-D(E/E_c)^{-3}-n_H\sigma(E)} \left[K_{PL} E^{-\alpha}+ \sum_i \frac{K_{G,i}}{\sqrt{2\pi}\sigma_i}e^{-\frac{(E-E_{0,i})^2}{2\sigma_i^2}} \right],& E\geqslant E_{c}; 
\end{cases} \\
\end{equation}
where $n_H$ is the column density of neutral hydrogen; $K_{PL}$ and $K_{G,i}$ 
are the normalization factor for the power-law component and the $i$th Gaussian component, 
respectively; $E_c$, $E_{0,i}$, and $\sigma_i$ are the threshold energy of the absorption edge, the center and the standard deviation of the $i$th Gaussian component, respectively; 
$D$ is the absorption depth at the threshold energy $E_c$; 
and $\alpha$ is the photon index of the power-law component in the model. The edge component at $\sim0.5$~keV is replaced by a Gaussian component (the $0$th component in Table~\ref{XMMspec}) for data taken on May 3 since the latter provides a better fit. We fix the column density of Galactic neutral hydrogen to $N_H \approx 1.9  \times10^{20} \;\text{cm}^{-2}$, which was measured by the Leiden/Argentine/Bonn (LAB) survey toward the direction of Mrk~421 \citep{Kalberla05}. It is worth noting that the best-fit power-law index hardly changes when we set $N_H$ free. 

The count rate measured by the EPN camera with a thin filter can be converted to flux using energy conversion factors (ECF, in units of $10^{11}$ cts cm$^2$ erg$^{-1}$), which depend on the filter, the photon index $\alpha$, the Galactic $n_H$ absorption, and the energy range \citep{Mateos09}. The flux $f$, in units of ergs cm$^{-2}$ s$^{-1}$, can be obtained by $f=rate/ECF$, where rate has the units of cts s$^{-1}$. 
A similar flux conversion factor is used for the OM UVM2 filter to convert each count at 2310 $\textup{\AA}$ to flux density $2.20\times10^{15} \text{ergs}\; \text{cm}^{-2}\;\text{s}^{-1} \; \textup{\AA}^{-1}$. 
A 2\% systematic uncertainty error was added to the OM light curve. 

The long-term {\it Swift}-XRT light curve is produced using an online analysis tool {\it The Swift-XRT data products generator} \footnote{\url{http://www.swift.ac.uk/user_objects/}} \citep{Evans09}. This tool is publicly available and can be used to produce {\it Swift}-XRT spectra, light curves, and images for a point source. 
A light curve of Mrk~421 was made from all {\it Swift}-XRT observations available from 2005 Mar 1 to 2014 Apr 30, integrated between 0.3 and 10 keV, with a fixed bin width of 50 s. We cut the first 150~s of each WT observation, during which it is possible that the satellite could still be settling, thus causing non-source-related deflections in the light curve. 

\subsection {Steward Observatory}

Regular optical observations of a sample of gamma-ray-bright blazars, including Mrk~421, have been carried out at Steward Observatory since the launch of the {\it Fermi} satellite~\citep{Smith09}, and these data are publicly accessible \footnote{\url{http://james.as.arizona.edu/~psmith/Fermi/}}. For the 2014 April-May MWL observing campaign, the SPOL optical, dual-beam spectropolarimeter \citep{Schmidt1992} was used at the Steward Observatory 1.54-m Kuiper Telescope on Mt. Bigelow, Arizona from April 25 to May 4 UTC. When the weather permitted, the usual observing frequency of one observation per night for Mrk~421 was increased to four per night after April 26 so that any rapid changes in linear polarization and optical flux could be better tracked. The spectropolarimeter was configured with a 600 l/mm diffraction grating that yields a dispersion of 4 $\textup{\AA}$ pixel$^{-1}$, spectral coverage from 4000-7550 $\textup{\AA}$, and resolution of $\sim16 \; \textup{\AA}$. The CCD detector is a thinned, anti-reflection coated $1200\times800$ STA device with a quantum efficiency of about 0.9 from 5000-7000 $\textup{\AA}$. All polarization observations of Mrk~421 were made with a 3"$\times$50" slit oriented so that its long (spatial) dimension is east-west on the sky and the CCD was binned by two pixels ($\sim$0.9") in the spatial direction. An observation of Mrk~421 typically consists of a 30-second exposure at all 16 positions of the $\lambda$/2-wave plate, properly sorted into four images with each image containing the two orthogonal polarized beams created by a Wollaston prism in the optical path. Extraction of the sky-subtracted spectra of Mrk~421 is done using a 3"$\times$9" aperture for all polarization observations to keep the contribution of the unpolarized starlight from its host galaxy as constant as possible for all measurements. Medians in the wavelength range of 5000-7000 $\textup{\AA}$ are taken of the resulting spectra of the linear Stokes parameters $q$ and $u$ and used to calculate the observed degree of polarization ($P$) and the position angle of the polarization on the sky ($\chi$). The instrumental polarization of SPOL has been consistently measured to be $\ll0.1$\% and is ignored. Likewise, Galactic interstellar polarization is negligible in the direction of Mrk~421 based on the amount of reddening estimated in this line of sight \citep[Av$\sim$0.042; ][]{Schlafly2011}. Given the lack of significant instrumental and Galactic interstellar polarization, the uncertainties in the measurements made by SPOL are dominated by photon statistics and typically $\sigma_p < 0.1$\% when the spectropolarimetry is binned by 2000 $\textup{\AA}$. The polarization position angle was calibrated during the campaign by observing the polarization standard stars Hiltner 960 and VI Cyg \#12 \citep{Schmidt1992a}.

Optical flux monitoring of Mrk~421 during this period was accomplished by using SPOL with a $7.6"\times50"$ slit when conditions were clear. As with the spectropolarimetry, the slit is oriented east-west on the sky and although the larger slit admits more host galaxy starlight, it minimizes slit losses as a function of wavelength. Differential photometry with ``Star 1'' \citep{Villata1998} was used to calibrate the V-band magnitude of Mrk~421 within a spectral extraction aperture of $7.6"\times9"$ . Generally, single 30-second exposure at a set wave plate position is obtained for both the blazar and the comparison star. A standard Johnson V filter bandpass transmission curve is multiplied to the extracted spectra and the instrumental fluxes for the objects are compared to derive the brightness of Mrk~421. This measurement is typically performed twice per visit to Mrk~421 to check the consistency of the photometry. The dominant source of uncertainty for the flux measurements is the precision of the V-band calibration of the comparison star (0.02 mag). The flux contribution of the host galaxy in R-band for a rectangular aperture of $7.6"\times9"$ centered at the Mrk~421 was estimated using the measurements in \citet{Nilsson2007}, and converted to V-band using an E galaxy template of age 11~Gyr at a redshift of 0.031 that gives a $V-R$ of 0.686 \citep{Fukugita1995}. 

\subsection {OVRO and CARMA}
Contemporaneous observations of Mrk~421 were taken with the Owens Valley Radio Observatory (OVRO) at 15~GHz \citep{Richards11} and the Combined Array for Research in Millimeter-Wave Astronomy (CARMA) at 95~GHz \citep{Bock06}. 

The OVRO 40~m telescope is equipped with a cryogenic, low-noise high electron mobility transistor amplifier with a 15.0~GHz center frequency and 3~GHz bandwidth. The two off-axis sky beams are Dicke-switched with the source alternating between the two beams, in order to remove the atmospheric and ground contamination. The receiver gain is calibrated using a temperature-stable diode noise source. 
The systematic uncertainty in the flux density scale is estimated to be approximately 5\%, and is not included in the error bars. More details of the reduction and calibration procedure can be found in \citet{Richards11}. 

The CARMA observations of Mrk~421 were made using the eight 3.5-m telescopes of the array with a central frequency of 95 GHz and a bandwidth of 7.5 GHz. 
The amplitude and phase gain were self-calibrated on Mrk 421. The absolute
flux was calibrated from a temporally nearby observation of the planets Mars, Neptune or Uranus, or the quasar 3C 273. 
The absolute systematic uncertainty is estimated to be approximately 10\%, and is not included in the error bars.
\begin{figure}[ht!]
  \centering
\includegraphics*[width=1.0\textwidth]{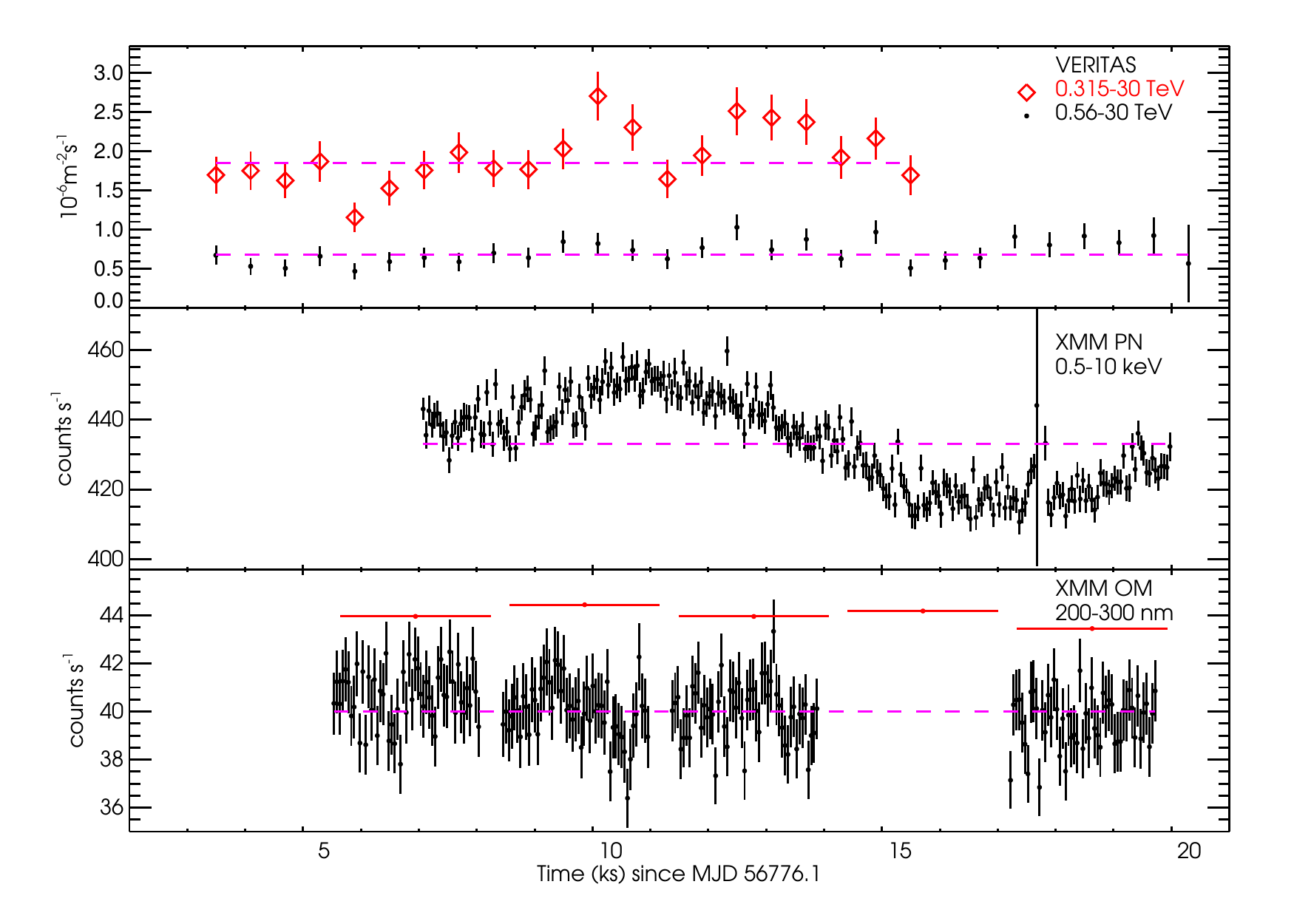}
\caption{{\it XMM-Newton} and VERITAS light curves of Mrk 421 from 2014 Apr 29 simultaneous ToO observations. Top panel: VERITAS flux light curves, integrated above the highest energy threshold of all runs on that night in 10-minute bins. Middle panel: {\it XMM}-EPN count rates between 0.5 and 10 keV in 50-s bins. Bottom panel: The black points are {\it XMM}-OM fast mode optical count rates between 200 and 300 nm in 50-s intervals, and the red points are OM image mode count rates binned by exposure.  }
                \label{0429LC}
\end{figure}
\begin{figure}[ht!]
\includegraphics*[width=1.0\textwidth]{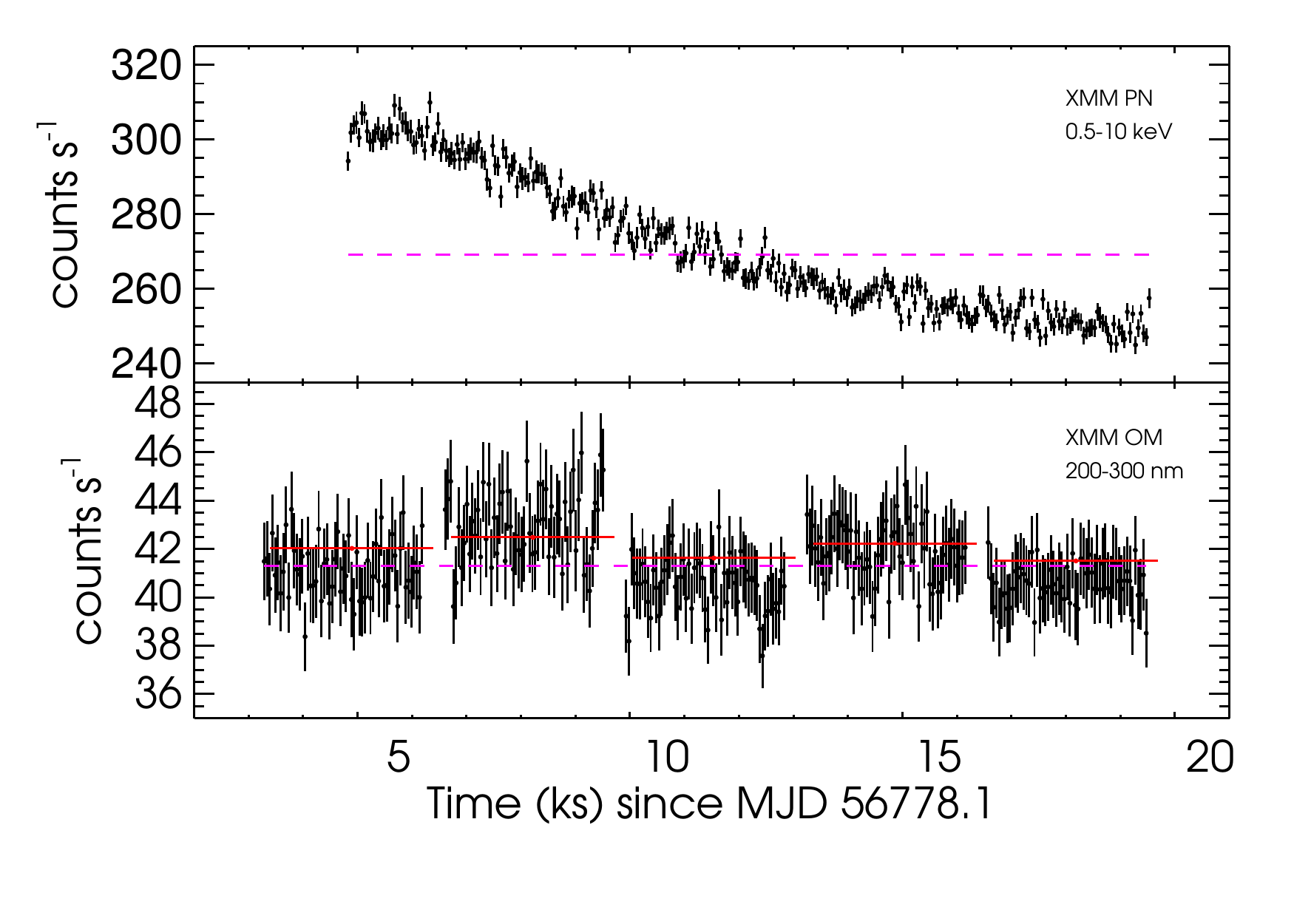}
\caption{{\it XMM-Newton} light curves of Mrk 421 from 2014 May 1 ToO observations. Top panel: {\it XMM}-EPN count rates between 0.5 and 10 keV in 50-s bins. Bottom panel: The black points are {\it XMM}-OM fast mode optical count rates between 200 and 300 nm in 50-s bins, and the red points are OM image mode count rates binned by exposure. Note that VERITAS data on May 1 are not shown because the data were taken under poor weather conditions. }
                \label{0501LC}
\end{figure}
\begin{figure}[ht!]
\includegraphics*[width=1.0\textwidth]{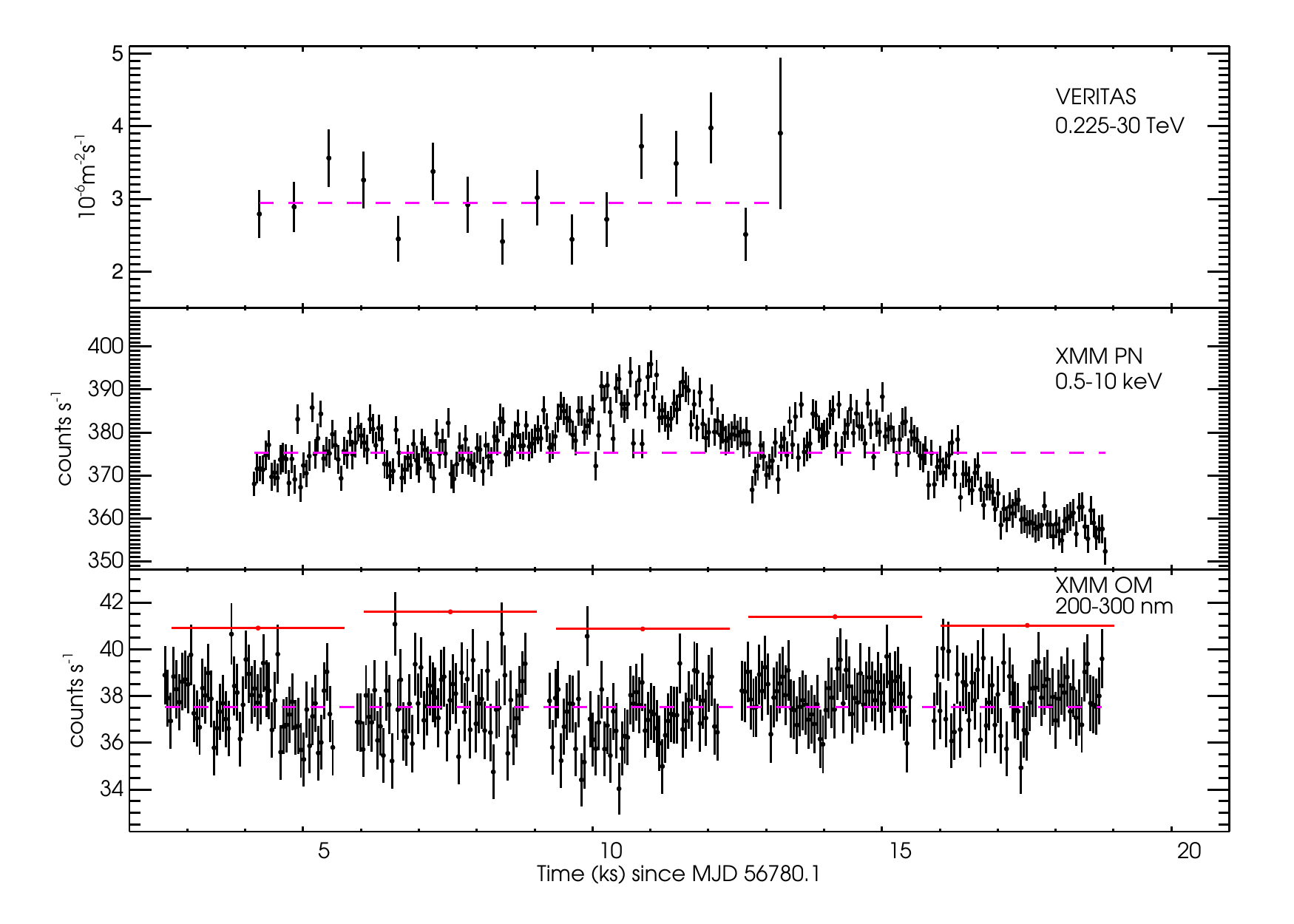}
\caption{{\it XMM-Newton} and VERITAS light curves of Mrk 421 from 2014 May 3 simultaneous ToO observations. Top panel: VERITAS flux light curves, integrated above the highest energy threshold of all runs on that night in 10-minute bins. Middle panel: {\it XMM}-EPN count rates between 0.5 and 10 keV in 50-s bins. Bottom panel: The black points are {\it XMM}-OM fast mode optical count rates between 200 and 300 nm in 50-s bins, and the red points are OM image mode count rates binned by exposure. }
                \label{0503LC}
\end{figure}
%
\section{Results}
\label{sec:res}
\subsection{Light curves}
\label{sec:XV_LC}
Figures~\ref{0429LC}, ~\ref{0501LC}, and~\ref{0503LC} show simultaneous light curves in VHE, X-ray, and UV bands. The VERITAS light curves are binned in 10-minute intervals, and shown with the integrated flux above the highest energy threshold among all observations taken on the corresponding night: 560~GeV on April 29 (315~GeV for the first $\sim$3.5 hr) and 225~GeV on May 3. The higher energy threshold on April 29 is a result of the larger zenith angle at which the source was observed ($\sim$48$^\circ$ during the last 30-min exposure of the night), which consequently leads to more distant shower maxima from the telescope. 
The X-ray light curves in the middle panels show the count rates measured by {\it XMM}-EPN between 0.5 and 10 keV, binned in 50~s intervals. The bottom panels show UV light curves constructed from the {\it XMM}-OM count rate using the UVM2 filter in both Image and Fast modes.

\begin{table}
  \centering
    \caption{ Reduced $\chi^2$ values for a constant fit to the light curves and the corresponding p-values for VERITAS light curves. }
    \begin{tabular}{ccccc} \\ \hline\hline
      Date & \multicolumn{2}{c}{VERITAS} & {\it XMM}-EPN & {\it XMM}-OM   \\
      & $\chi^2_\text{red}$ & p-value & $\chi^2_\text{red}$ & $\chi^2_\text{red}$ \\ \hline 
      \multirow{2}{*}{2014-04-29} & 2.1 ($>315$ GeV) & 0.003 & \multirow{2}{*}{11.1} & \multirow{2}{*}{0.9}   \\
        & 1.2 ($>560$ GeV) & 0.2 & &    \\
      2014-05-01 & - & - & 48.0  & 0.9  \\ 
      2014-05-03 &  1.6 & 0.07 & 7.0 & 0.9 \\ \hline 
      \end{tabular} \\
      \label{chi2}
\end{table}

The average VERITAS integral fluxes above 0.4 TeV are $(1.27 \pm 0.03)\times10^{-6} \;\text{photons} \;\text{m}^{-2} \text{s}^{-1}$ on Apr~29 and $(1.10 \pm 0.04)\times10^{-6} \;\text{photons} \;\text{m}^{-2} \text{s}^{-1}$ on May~3. As shown in Table~\ref{chi2}, a constant fit to the X-ray light curves yields large reduced $\chi^2$ values (corresponding p-values $<1\times10^{-5}$, thus rejecting the hypothesis of constant flux), implying the presence of intra-night variability. The corresponding p-values in the VHE band, of 0.003 and 0.07, imply marginally-significant intra-night variabilities in the VERITAS light curves $>315$ GeV on Apr 29 and $>225$ GeV on May 3. %

Another quantity that describes the relative amount of variability is the fractional variability amplitude. Following the descriptions in \citet{Vaughan03} and \citet{Poutanen08}, the fractional variability $F_{var}$ and its error $\sigma_{F_{var}}$ are calculated as 
\[
F_{var}= \sqrt{\frac{S^2-\langle{\sigma_{err}^2}\rangle}{\langle F \rangle^2}}
\]
and 
\[
\sigma_{F_{var}}= \sqrt{F_{var}^2+\sqrt{\frac{2 \langle \sigma_{err}^2\rangle^2}{N\langle F\rangle^4} + \frac{ 4 \langle{\sigma_{err}^2}\rangle F_{var}^2}{N \langle F \rangle^2}}}-F_{var},
\]
where $S$ is the standard deviation of the N flux measurements, $\langle{\sigma_{err}^2}\rangle$ is the mean squared error of these flux measurements, and $\langle F \rangle$ is the mean flux. 

\begin{figure}[ht!]
  \centering
  \includegraphics[width=0.8\textwidth]{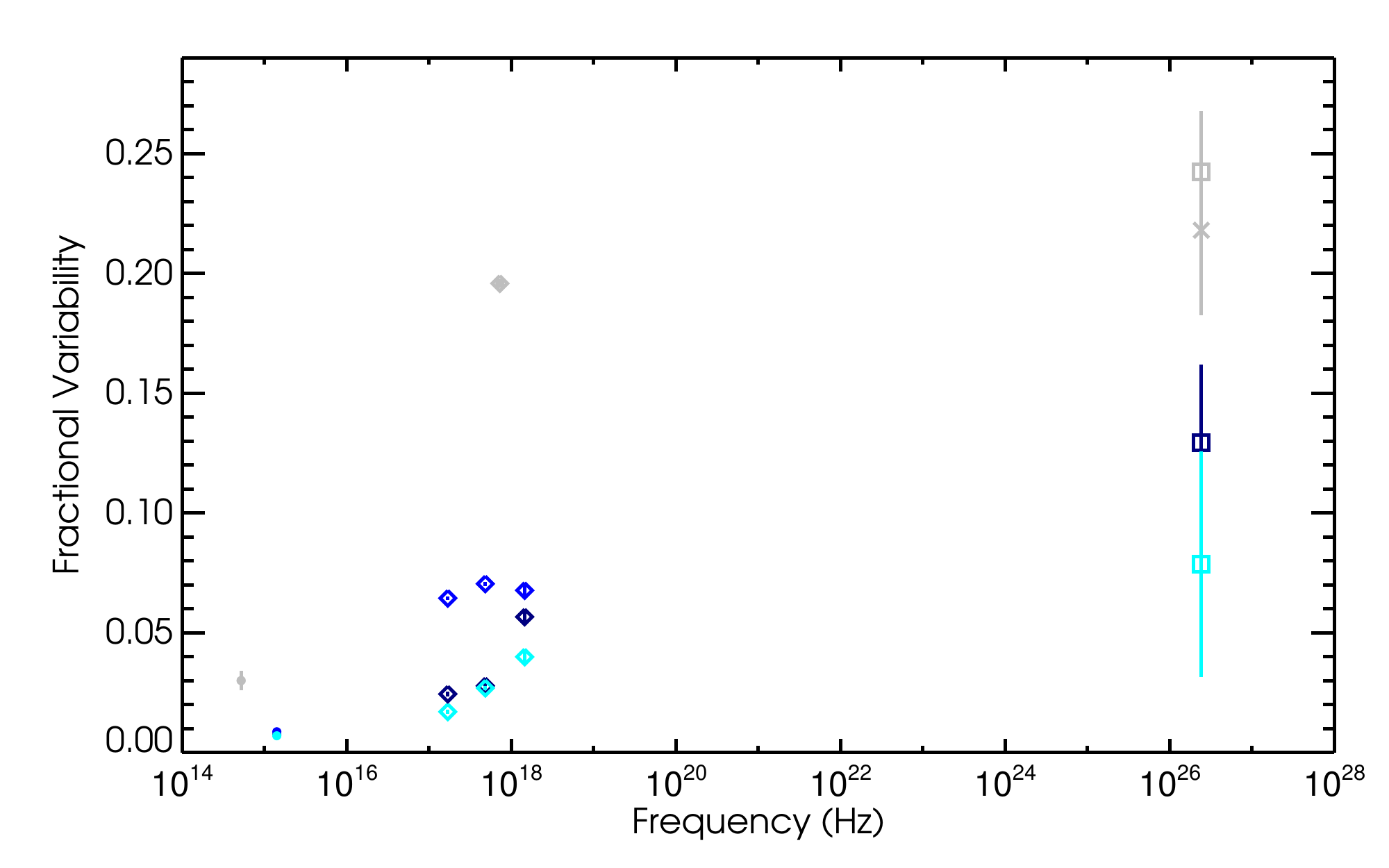}
  \caption{Fractional variability of VHE and X-ray light curves of Mrk 421 from the three simultaneous ToO observations in 2014. Open squares are calculated from VERITAS light curves with 600-s time bins and $\sim$15-ks duration on the two nights under good weather conditions, and open diamonds are from {\it XMM}-EPN light curves with 50-s time bins and $\sim$15-ks duration. The results from three energy intervals (0.5-1 keV, 1-3 keV, and 3-10 keV) in the X-ray band are shown. Navy points represent the measurements on April 29, blue points for May 1, cyan ones for May 3, and gray ones for the duration of one week. The gray open square is from the VERITAS one-week flux measurements, and the gray cross is from the MAGIC one-week flux measurements. Both VHE fluxes are above 560 GeV with a 30-minute bin width. The gray diamonds are calculated from the XRT light curve with a 50-s bin width and a one-week duration, and the gray filled circles are from the Steward Observatory light curve sampled at intervals of a few hours with one-week duration. }
  \label{NVA}
\end{figure}
The fractional variability results from the simultaneous {\it XMM-Newton} and VERITAS data, as well as from contemporaneous MWL data, are shown in Figure~\ref{NVA}. The VHE fractional variability has been computed using the flux light curves integrated above 315~GeV for data from the first 3.5 hr on April 29, and above 225~GeV on May 3. The VERITAS fractional variability is $\sim$$13\%\pm3\%$ on Apr 29, and $\sim8\%\pm5\%$ on May 3. %

The fractional variability of the X-ray flux is low, but significantly above zero in all three energy intervals 0.5-1 keV, 1-3 keV, and 3-10 keV. 
Comparing these three X-ray energy intervals, a higher fractional variability is observed at higher frequencies on April 29 (navy open diamonds) and on May 3 (cyan open diamonds), in agreement with previous results \citep[e.g.][]{Blazejowski05}. 
This may be explained as the manifestation of a different synchrotron-cooling time at different energies, $t_{cool} \propto E^{-1/2}$. In the slow-cooling regime (the cooling time $t_{cool}$ is longer than the dynamic timescale $R/c$), the cooling time is shorter for higher-energy particles, leading to a faster variability in radiation at higher energies. Therefore, more variation at higher energies is observed compared to lower energies on the same timescales, which directly leads to a higher fractional variability for higher energy emissions. 
However, the same trend is less obvious on May 1 (blue open diamonds), when only X-ray data are available. 
\begin{figure}[!h]
  \centering
   \includegraphics[height=0.8\textheight]{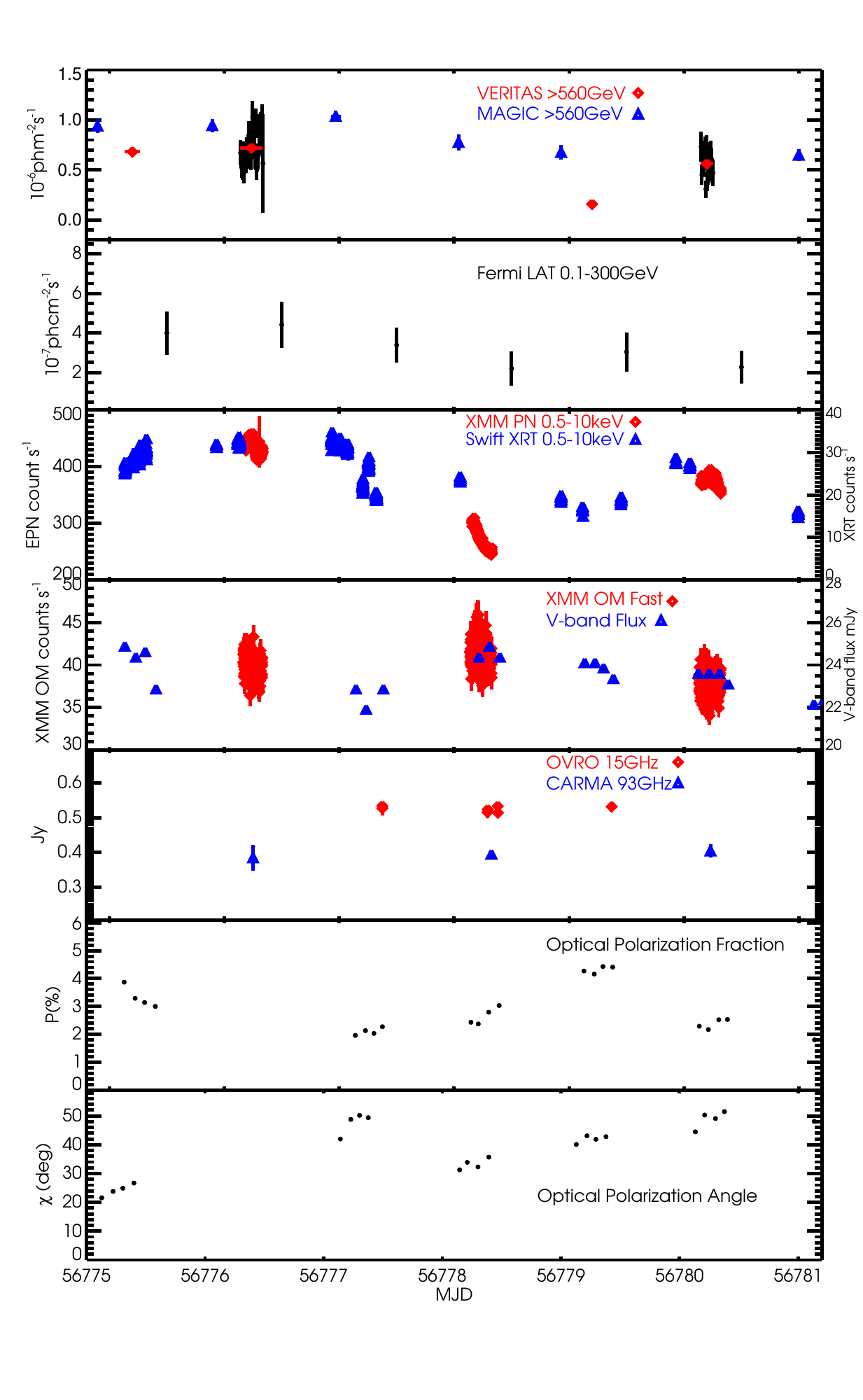}
  \caption{MWL light curves between Apr 28 and May 4. See text for details of the light curve in each panel. }
  \label{MWLLC}
\end{figure}

Contemporaneous MWL observations often provide valuable information about the activity of the source, e.g. abrupt changes in the radio and optical polarizations would reveal the emergence of a compact region that may be connected to a flaring event \citep[see e.g.][]{Arlen13}. MWL light curves of Mrk~421 from MJD 56775 to 56781 are shown in Figure~\ref{MWLLC}. The TeV light curve, measured by MAGIC above 560~GeV on each night before the {\it XMM-Newton} observations, is shown in blue in the top panel. 
The fractional variabilities of the VHE flux for a one-week duration with a 30-min bin width are measured by both the VERITAS and the MAGIC are shown as the gray open square and the gray cross in Figure~\ref{NVA}, respectively. 
Only statistical uncertainties are taken into account in the calculation of the VHE fractional variability. As mentioned in Section~\ref{subsec:VHE}, the night-to-night systematic uncertainty in the VHE flux measurements from MAGIC in estimated to be $\sim$11\%, with the primary contribution to which is the fluctuation of the atmospheric transmission \citep[see][for further details]{aleksic2016}. 
We follow a similar approach and estimate the systematic uncertainty in the VHE flux measured by VERITAS with 10-min intervals to be less than $\sim$10\% using observations of the Crab Nebula under similar conditions as the observations of Mrk~421 in this work. 
Adding a 10\% systematic uncertainty in quadrature to the statistical uncertainty of the VERITAS-measured VHE flux reduces its one-week $F_{var}$ value from $24.2\%\pm2.5\%$ to $21.9\%\pm3.6\%$, and the $F_{var}$ value on Apr 29 from $\sim$$13\%\pm3\%$ to $\sim$$8\%\pm5\%$, and the $F_{var}$ value $\sim8\%\pm5\%$ on May 3 should be considered as an upper limit. 

The daily {\it Fermi}-LAT light curve (shown in the second panel) does not suggest any significant variability, although there might be a slight drop in GeV flux after May 1. X-ray count rates from the {\it Swift}-XRT are also shown, together with those measured by {\it XMM}-EPN in the third panel. 
The {\it Swift}-XRT results fill the gaps between the three {\it XMM-Newton} observations, and also show significant variability, illustrated by the fractional variability of $\sim$20\% computed from XRT data between Apr~28 and May~4 shown as the gray diamond in Figure~\ref{NVA}. Note that fractional variabilities of $\sim$20\% to $\sim$40\% in the X-ray and $\sim$30\% in VHE during typical non-flaring states were found by \citep{aleksic2015}, and much higher values, of $\sim$40\% to $>$60\% in both X-ray and VHE during flaring states were found by \citet{Blazejowski05} and Aleksi{\'c} (2016 in prep). It is worth noting that fractional variability depends on the bin width, the sampling frequency, and the duration of the light curve (see discussion section), which makes it more difficult to compare the $F_{var}$ values measured using different light curves. 
%

%
The optical photometric variability of Mrk~421 from April 25 to May 4 is mild (see the fourth panel of Figure~\ref{MWLLC}). 
Mrk~421 varied from 12.8 to 12.9 in V band, with the maximum optical flux observed on MJD 56775. Over a 24-hour period, the object showed a maximum $\Delta V \sim 0.1$ mag and intra-night variability was generally $<0.05$ mag. During this period, Mrk~421 was close to the middle of the range of optical brightness it has shown since 2008 (V$\sim$11.9-13.6).
The source does not exhibit strong variability in the UV band, nor at 15~GHz or 95~GHz (see the fourth and fifth panel of Figure~\ref{MWLLC}). 

In contrast to the flux variations, the optical polarization of Mrk~421 showed more pronounced variability during the dates shown. 
The observed polarization fraction $P$ peaked at MJD 56779 (4-5\%) with minima of $P\sim$2\% two days preceding and one day after the polarization maximum (see the sixth panel of Figure~\ref{MWLLC}). 
The polarization peak reaches only about half of the highest polarization levels observed for this object (10-13\%) since 2008. From 2008-2015, the Steward Observatory blazar data archive identify several periods when the polarization of Mrk~421 is $<1$\%. Like many blazars, the full range of values for the polarization position angle is exhibited by Mrk~421 over a time scale of several years. 
During the dates shown, the polarization angle $\chi$ varied between 20$^\circ$ and 55$^\circ$ with rotations of nearly 20$^\circ$ observed in a day (see the bottom panel of Figure~\ref{MWLLC}). 
Intra-night variations as large as $\sim10^\circ$ were observed on time scales as short as two hours. During the epoch of the campaign, $\chi$ is roughly orthogonal to the position angle of the 43 GHz VLBI jet ($\sim-35^\circ$) \footnote{\url{https://www.bu.edu/blazars/VLBAproject.html}}, implying that the magnetic field within the region emitting the polarized optical continuum is more-or-less aligned with the jet.

%

\subsection{Cross-band flux correlation}
\label{sec:XVcorr}
\begin{figure}[!h]
  \centering
  \includegraphics[width=0.75\textwidth]{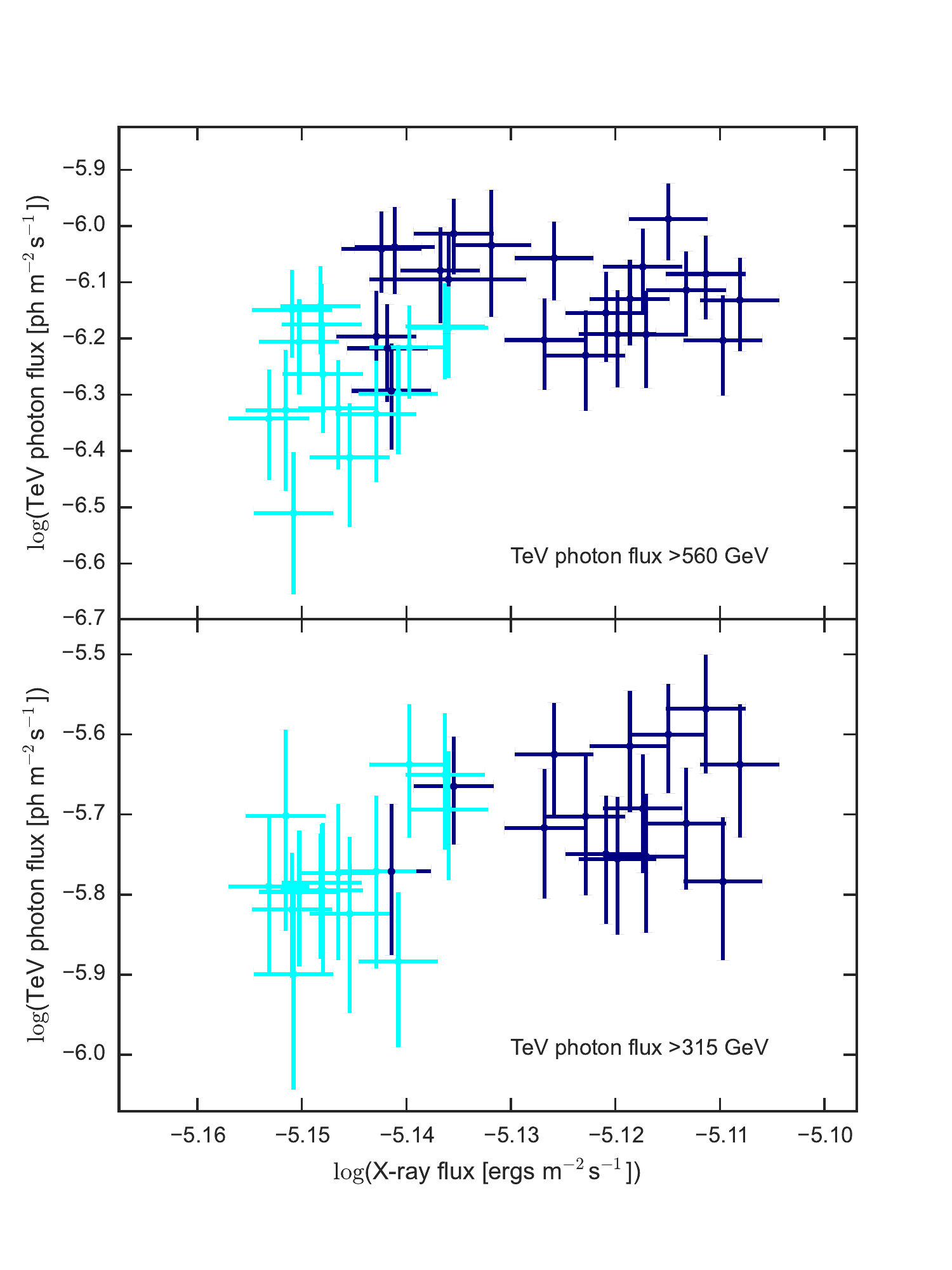}
  \caption{TeV photon flux versus X-ray energy flux from the simultaneous observations on 2014 April 29 (shown in navy) and May 3 (shown in cyan). The VHE fluxes are measured by VERITAS integrated above 560~GeV (top panel) and 315~GeV (bottom panel); X-ray energy flux values are converted from the {\it XMM}-EPN count rates using ECFs based on the best-fit photon index and neutral hydrogen density of each night. Both X-ray and TeV data are binned in 10-minute intervals. }
  \label{XVCorr}
\end{figure}

We plot the VHE flux against X-ray count rate in Figure~\ref{XVCorr}. There is no significant evidence for correlation. The Pearson correlation coefficient between X-ray flux and TeV gamma-ray flux $>560$ GeV is 0.48 (the 90\% confidence interval is 0.24-0.67), and that between X-ray flux and TeV gamma-ray flux $>315$~GeV is 0.60 (the 90\% confidence interval is 0.35-0.76). The values of the correlation coefficients only suggest a moderate positive correlation, without considering the uncertainties on the measurements. 
Therefore, we focus on the flux correlations between 
three X-ray bands, and between two TeV gamma-ray bands, respectively. 
\subsubsection{Hard/soft X-ray correlation}
\label{sec:X_crossband}
\begin{figure}[htp]
  \centering
      \begin{tabular}{ll} 
  \includegraphics[height=0.5\textheight]{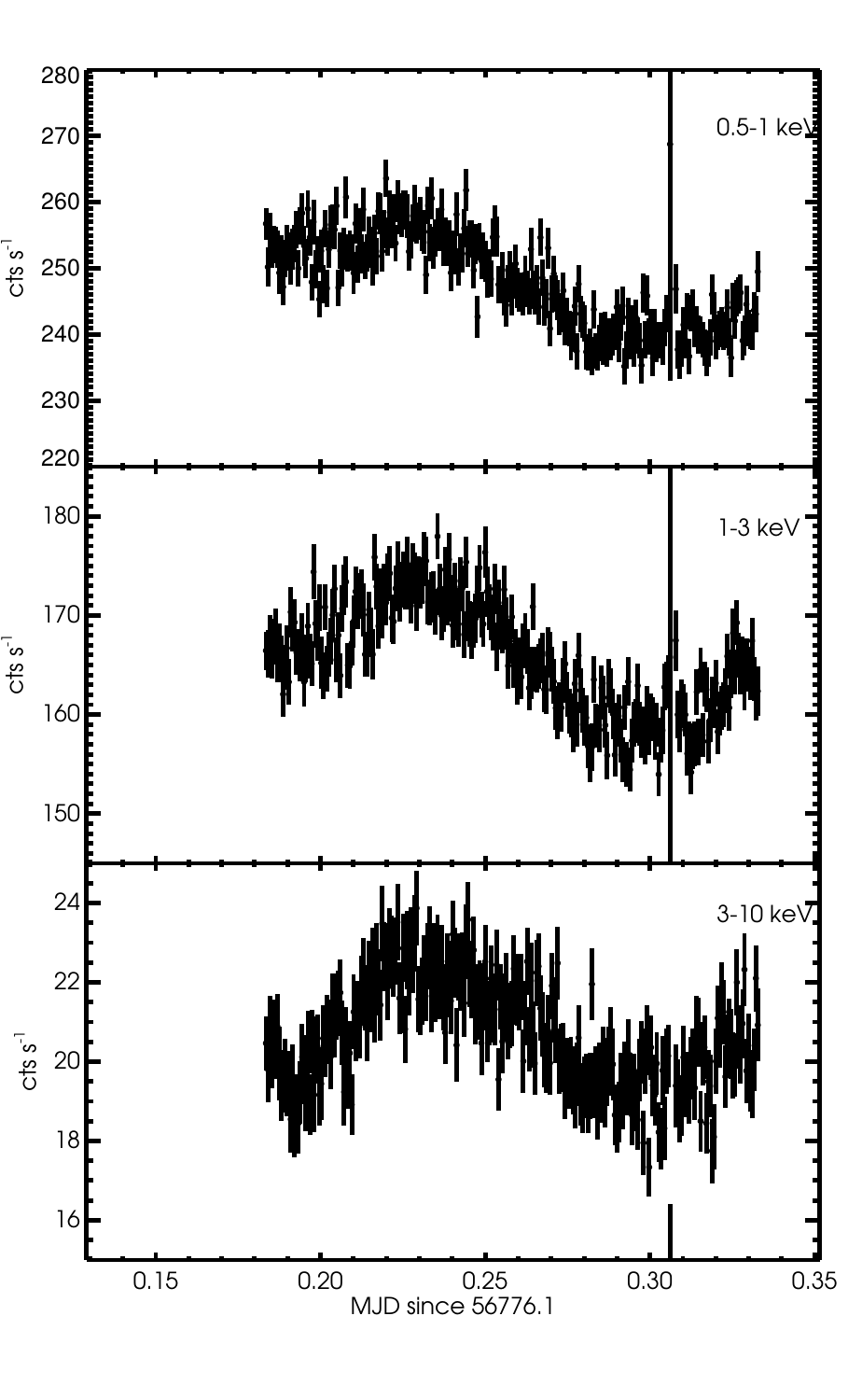} & 
    \includegraphics[height=0.5\textheight]{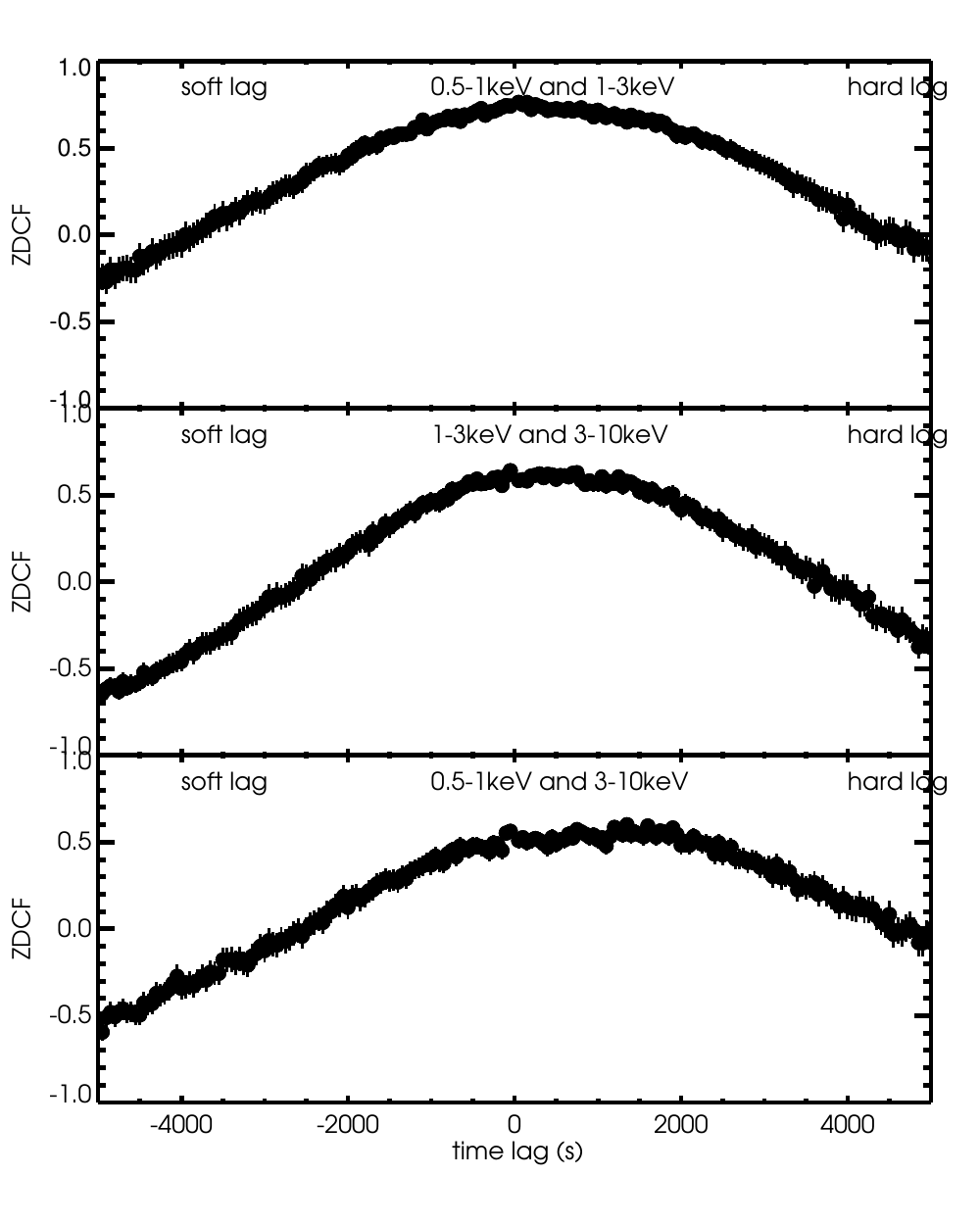}\\
      \end{tabular} \\ 
  \caption{ Left panel: Light curves of Mrk~421 observed with {\it XMM-Newton}-EPN on 2014 Apr~29. Count rates binned in 50~s time intervals in three energy bands, 0.5-1~keV, 1-3~keV, and 3-10~keV, are shown from top to bottom panel, respectively. Right panel: the ZDCF between these three X-ray bands. Positive lag values indicate ``hard lag''. } 
    \label{X_crossband0429}
\end{figure}

\begin{figure}[htp]
  \centering
      \begin{tabular}{ll} 
  \includegraphics[height=0.5\textheight]{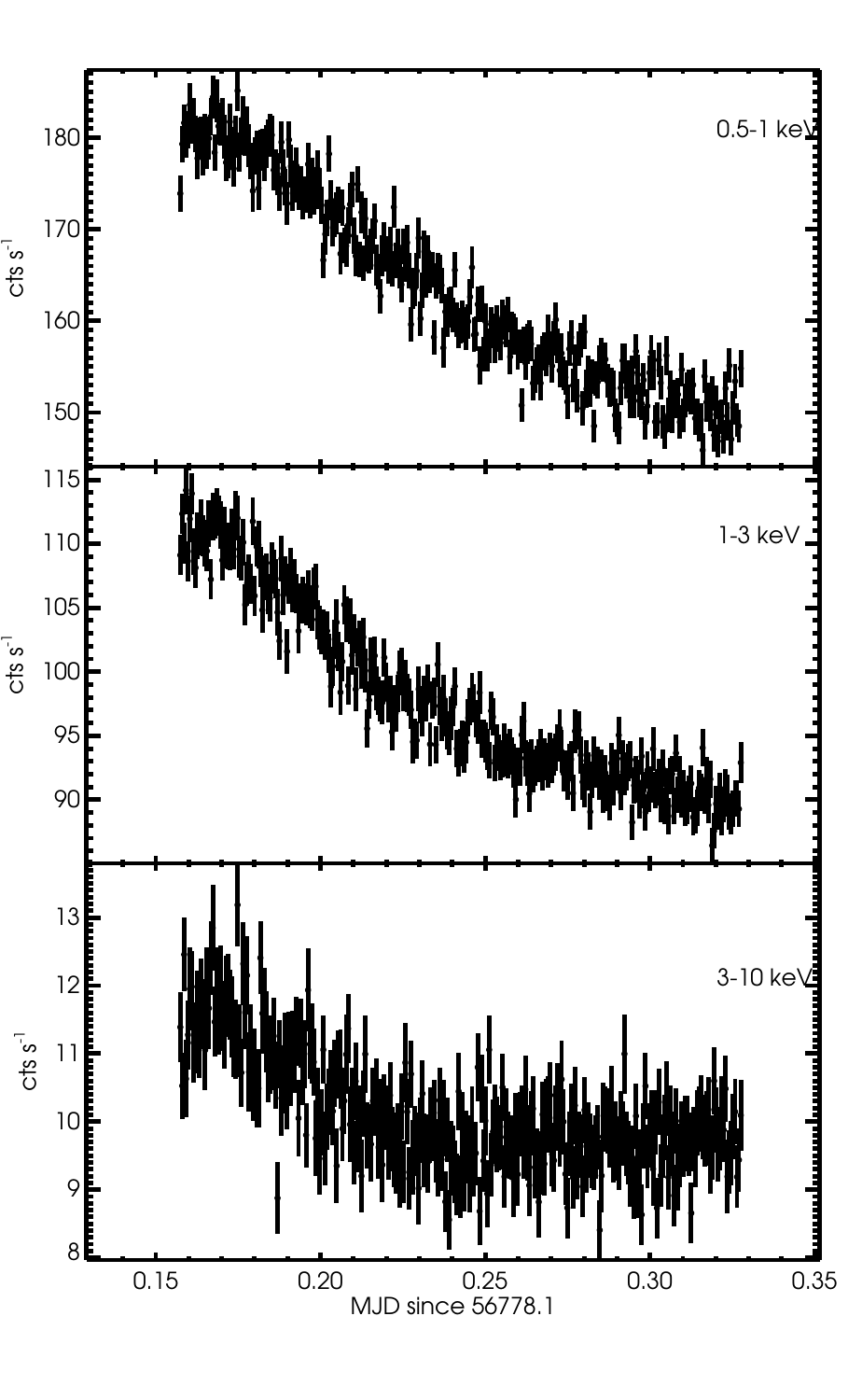} &
    \includegraphics[height=0.5\textheight]{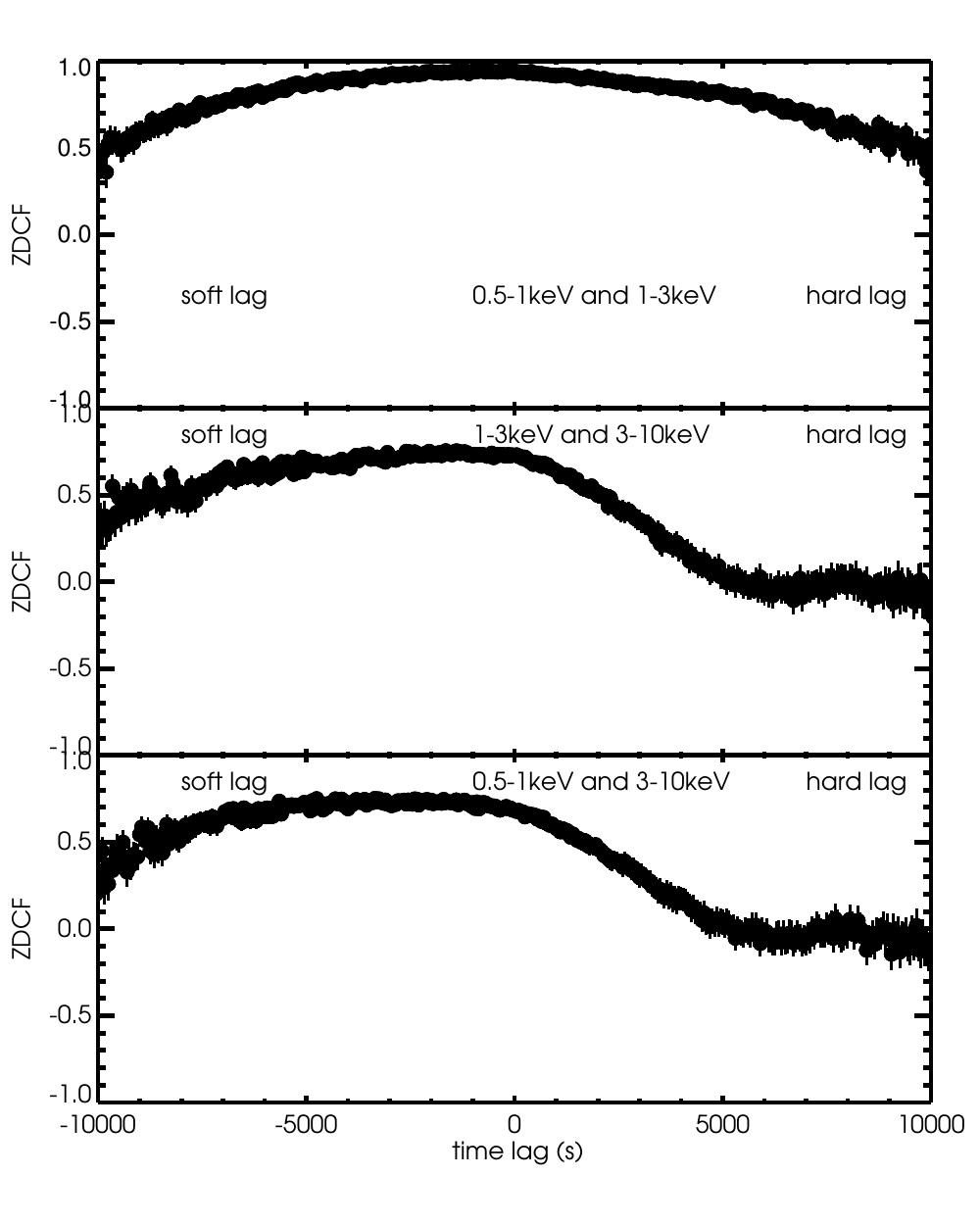}\\
      \end{tabular} \\ 
  \caption{ Left panel: Light curves of Mrk~421 observed with {\it XMM-Newton}-EPN on 2014 May~1. Count rates binned in 50~s time intervals in three energy bands, 0.5-1~keV, 1-3~keV, and 3-10~keV, are shown from top to bottom panel, respectively. Right panel: the ZDCF between these three X-ray bands. Positive lag values indicate ``hard lag''. } 
    \label{X_crossband0501}
\end{figure}

\begin{figure}[htp]
  \centering
      \begin{tabular}{ll} 
 \includegraphics[height=0.5\textheight]{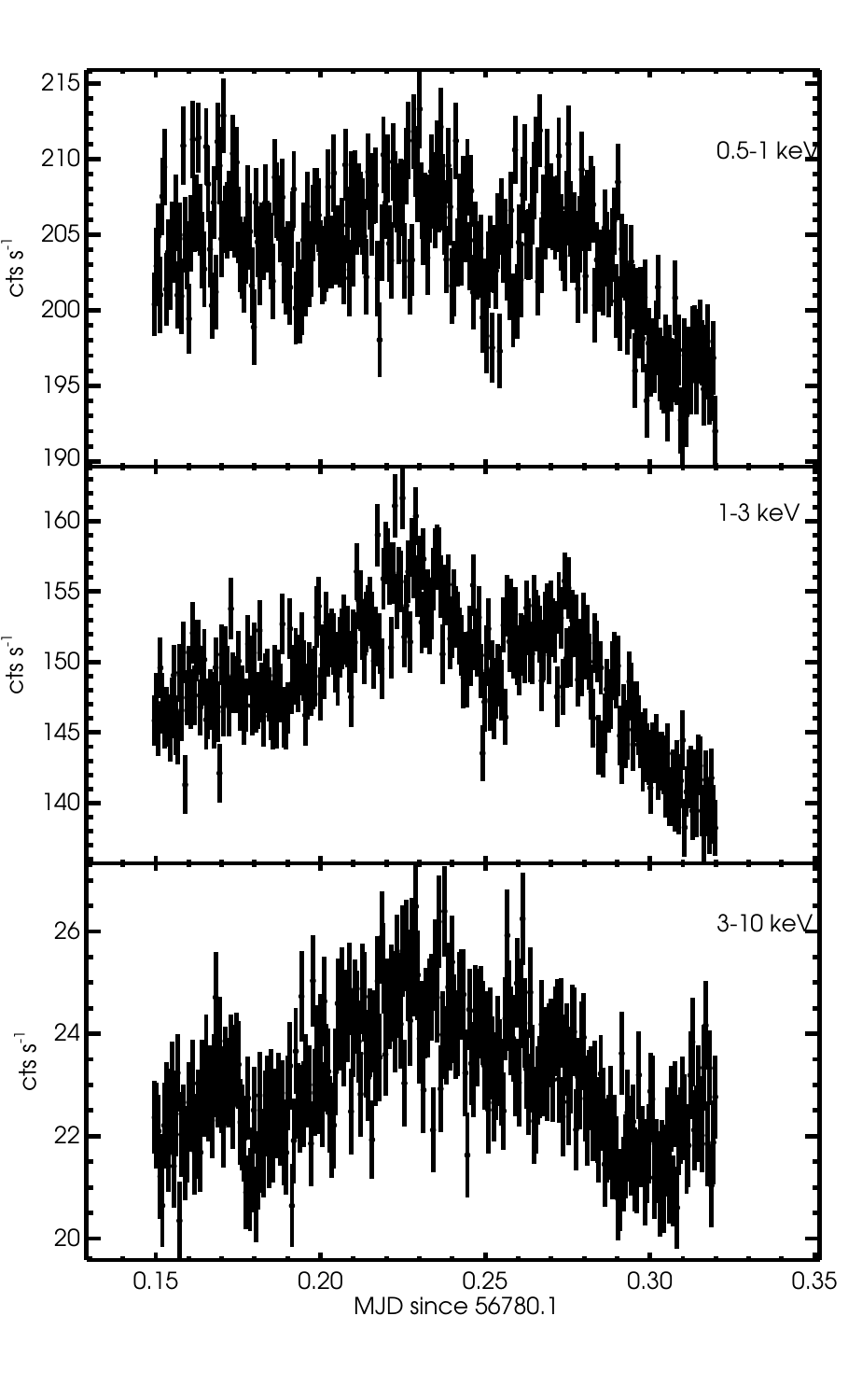} &
 \includegraphics[height=0.5\textheight]{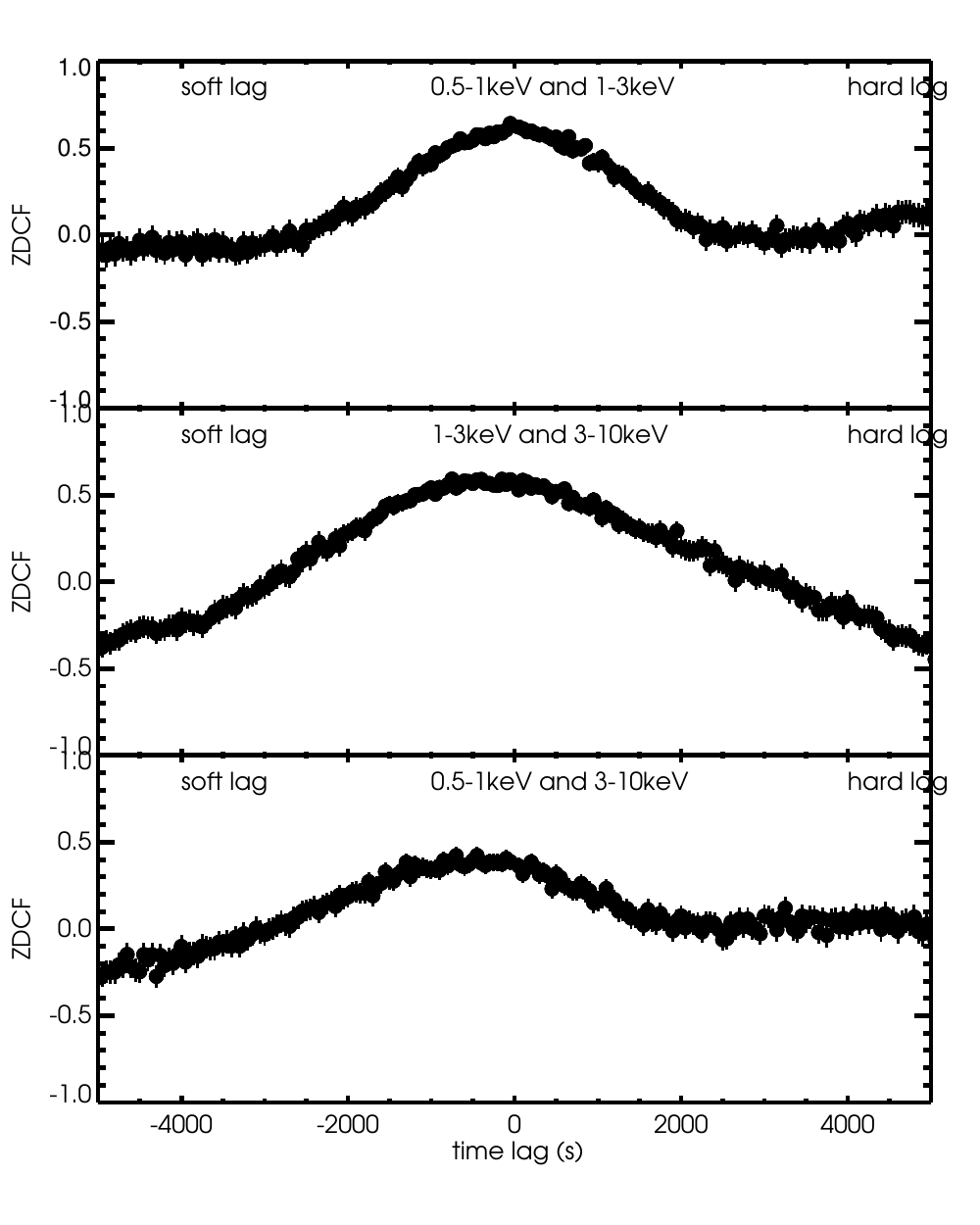}\\
       \end{tabular} \\
  \caption{ Left panel: Light curves of Mrk~421 observed with {\it XMM-Newton}-EPN on 2014 May~3. Count rates binned in 50~s time intervals in three energy bands, 0.5-1~keV, 1-3~keV, and 3-10~keV, are shown from top to bottom panel, respectively. Right panel: the ZDCF between these three X-ray bands. Positive lag values indicate ``hard lag''. } 
    \label{X_crossband0503}
\end{figure}

We further divide {\it XMM}-EPN X-ray light curves into three energy bands, 0.5-1~keV, 1-3~keV, and 3-10~keV (as shown in the left panels in Figures~\ref{X_crossband0429}, \ref{X_crossband0501}, and~\ref{X_crossband0503}). 
Z-transformed discrete correlation functions (ZDCFs) between these soft- and hard-X-ray light curves are calculated using a publicly-available code, \texttt{ZDCF v2.2} developed by \citet{Alexander13}, as shown in the right panels in Figures~\ref{X_crossband0429}, \ref{X_crossband0501}, and~\ref{X_crossband0503}. At least 11 pairs of light curve points in each time delay bin are required to calculate the ZDCF, zero lag is not omitted, and 1000 Monte Carlo runs were used to estimate the measurement error, in addition to the error calculated in the z-space. 
From the ZDCFs, the corresponding time lags are calculated using \texttt{PLIKE v4.0} also developed by \citet{Alexander13}. 
No significant evidence is found for any leads or lags between the soft-X-ray band relative to the hard-X-ray band. 

\subsubsection{Gamma-ray intraband correlation}
\label{sec:V_crossband}
The cross-correlations between light curves of blazars at TeV energies are particularly interesting, not only because they can provide insight to the particle acceleration and radiation, but also due to their potential to test Lorentz-invariance violation, which is a manifestation of an energy-dependent speed of light at the Planck scale predicted by foamy structures of space time in certain quantum theories \citep[e.g.][]{Aharonian08,Albert2008,Zitzer13}. 

\begin{figure}[!h]
  \centering
        \includegraphics[width=0.75\textwidth]{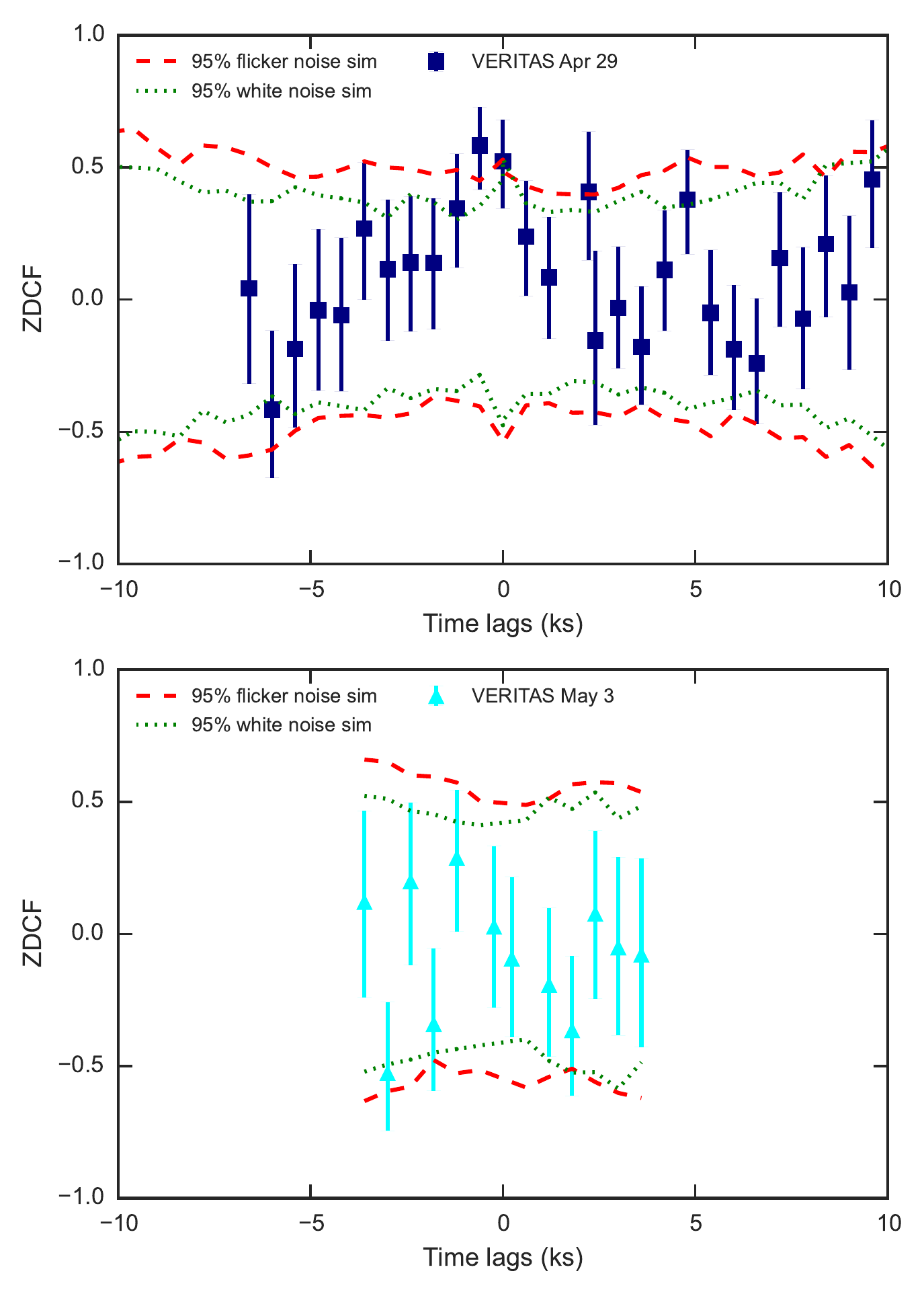}
    \caption{VERITAS ZDCFs between light curves integrated below and above 560~GeV of Mrk 421 on 2014 Apr~29 and May~3. The 95\% confidence region from flicker noise and white noise simulations are shown as red dashed lines and green dotted lines, respectively. }
    \label{VZDCF}
\end{figure}

In a similar fashion as done for the X-ray data (as described in Section~\ref{sec:X_crossband}), we divide the gamma-ray light curves into two bands, and compute ZDCFs and time lags as shown in 
Figure~\ref{VZDCF}. The chosen bands are 315-560~GeV and 560~GeV-30~TeV on Apr 29, and 225-560~GeV and 560~GeV-30~TeV on May 3, so that the event rates are comparable in the higher- and the lower-energy bins. 
ZDCFs are calculated using light curves binned by 10 minutes and 4 minutes, respectively. 
The 1-$\sigma$ confidence interval of the time lag of maximum likelihood is calculated between -2000~s and 2000~s using \texttt{plike\_v4.0}. 
To understand the ZDCFs produced by random noise, we simulate flicker-noise (whose power spectral density distribution is proportional to $1/f$) and Gaussian white noise with 10-min bin width and similar duration as the data. The 95$\%$ confidence regions calculated from ZDCF values between 200 pairs of simulated light curves are plotted in Figure~\ref{VZDCF}, along with ZDCFs calculated from 10-min binned VERITAS light curves above and below 560 GeV. 
No evidence for time leads or lags is present in the gamma-ray data; however, given the lack of a strong detection of variability, this lack of evidence for any leads/lags is not surprising since the sensitivity to such leads and lags is dependent on the amplitude of the detected variability in these data. 
%

\subsection{Hardness flux correlation and spectral hysteresis}
\label{XhystSec}
Besides the possibility of time lags at different energies, the spectral evolution during blazar flares is also informative. 
A general trend, that the spectrum is harder when the flux is higher, has been observed in blazars in both the X-ray and gamma-ray bands \citep[e.g.][]{Fossati08,Acciari11M4flare08}. 
Several possibilities can lead to such a trend if the spectrum is being measured at the high-energy end of the synchrotron and inverse-Compton SED peaks. For example, this could result from an increase in the maximum electron energy, or a hardening in the electron energy distribution. If, however, the X-ray and gamma-ray observations are sampling the emission near the two peaks of the SED, this ``harder-when-brighter'' effect could also be the result of an increase of the SED peak frequency, which could arise from an increase in magnetic field strength or Doppler factor. 

As well as the ``harder-when-brighter'' trend in the hardness-flux relation, 
competition between the acceleration and cooling timescales can lead to spectral hysteresis, since the spectral hardness differs on the rising and falling edges of flares \citep[e.g.][]{KRM98,Li00,Sato08}. 
If one plots the hardness-flux relation so that the spectrum is harder in the positive $y$-axis direction, and the flux is higher in the positive $x$-axis direction, the spectral hysteresis can be seen as a ``loop'' pattern.  
Since the spectral hysteresis is driven by the same timescales that determine the time lags at different energies, the direction of the hysteresis loop should be consistent with the sign of the time lag. Specifically, a ``hard lag'' should correspond to counter-clockwise hysteresis loops (see Section~\ref{sec:X_crossband}), while a ``soft lag'' will lead to clockwise hysteresis loops. Therefore, the hardness-flux plot offers an alternative view of the timescales in the system, and can be compared with the time-lag studies presented in the previous section. 
\begin{table}[!h]
  \centering
    \caption{ {\it XMM-Newton}-EPN spectral fit results and ECFs using the absorbed-power-law model plus three instrumental features described in Equation~\ref{XMMmodel}. Note that the fits are insensitive to the width $\sigma_{i}$  of the Gaussian component, therefore the equivalent width (EW) values are shown instead. The instrumental feature around 0.5 keV is fitted as an edge component on MJD 56776 and 56778, but as an Gaussian component on 56780. }
          \begin{tabular}{ccccc} \\ \hline\hline
       Parameter           	& Unit     			      	& Value on 56776         	& Value on 56778	     	& Value on 56780 	\\ \hline 
        $E_c$                 	& keV     			      	& $ 0.540\pm0.003 $  	&   $ 0.530\pm0.003 $  	&  -				\\
        $D$                     	&   				      	& $ 0.118\pm0.005 $        	& $ 0.146\pm0.009 $    	& - 				\\
        $E_{0,0}$          	& keV                                  	&   - 				     	&  -  				 	& $ 0.48\pm0.04 $ 	\\
        $EW_{0}$    	& keV  	                         	&   - 				     	&  -   					& $<0.11$			 	\\
        $K_{G,0}$         	&     $10^{-3}$       	      	&   - 		    		     	&  - 					&  $ 0.022 \pm 0.015 $ 		\\      
        $n_H$                	& $10^{20}$ cm$^{-2}$  	& $ 1.9 $              		& $ 1.9 $ 				&  $ 1.9 $ 			\\
        $\alpha$            	&  				      	& $2.649 \pm 0.002 $       	& $ 2.817\pm0.004 $     	& $ 2.450\pm 0.002$ \\
        $K_{PL}$           	&  			               	& $ 0.2714\pm 0.0004$ 	& $ 0.1711\pm0.0004 $  	& $ 0.2251\pm0.0002 $ \\
        $E_{0,1}$          	& keV                                  	& $ 1.88\pm0.02 $  	     	& $ 1.88\pm0.02 $  	 	& $ 1.87\pm0.03 $ 	\\
        $EW_{1}$    	& eV  	                         	&   $4 \pm 2$ 			&  $4 \pm 3 $ 			& $2 \pm 2$		\\
        $K_{G,1}$         	&     $10^{-4}$       	      	& $2.2 \pm 0.4 $ 	     	&  $ 1.1\pm0.4 $ 		&  $ 0.9\pm0.5 $ 	\\
        $E_{0,2}$          	& keV  		               	& $ 2.26\pm 0.01$       	&  $ 2.25\pm0.01 $ 		&  $2.26 \pm 0.01$ 	\\
        $EW_{2}$    	& eV  	                         	&   $26 \pm 3$			&  $19  \pm 3$ 			& $19 \pm 7$	 	\\
        $K_{G,2}$         	&      $10^{-4}$                  	& $ 8.1\pm 0.4 $  	      	& $ 3.3\pm0.4 $  		& $ 5.9\pm0.6 $	 \\
        Reduced $\chi^{2}$ &                                 	&        1.6 		      		&  1.3  				& 2.3 			\\	\hline				          
       \end{tabular} \\      
      \label{XMMspec}
\end{table}
\begin{figure}[ht]
  \centering
\subfloat[2014 April 29]{%
\includegraphics*[width=0.4\textwidth]{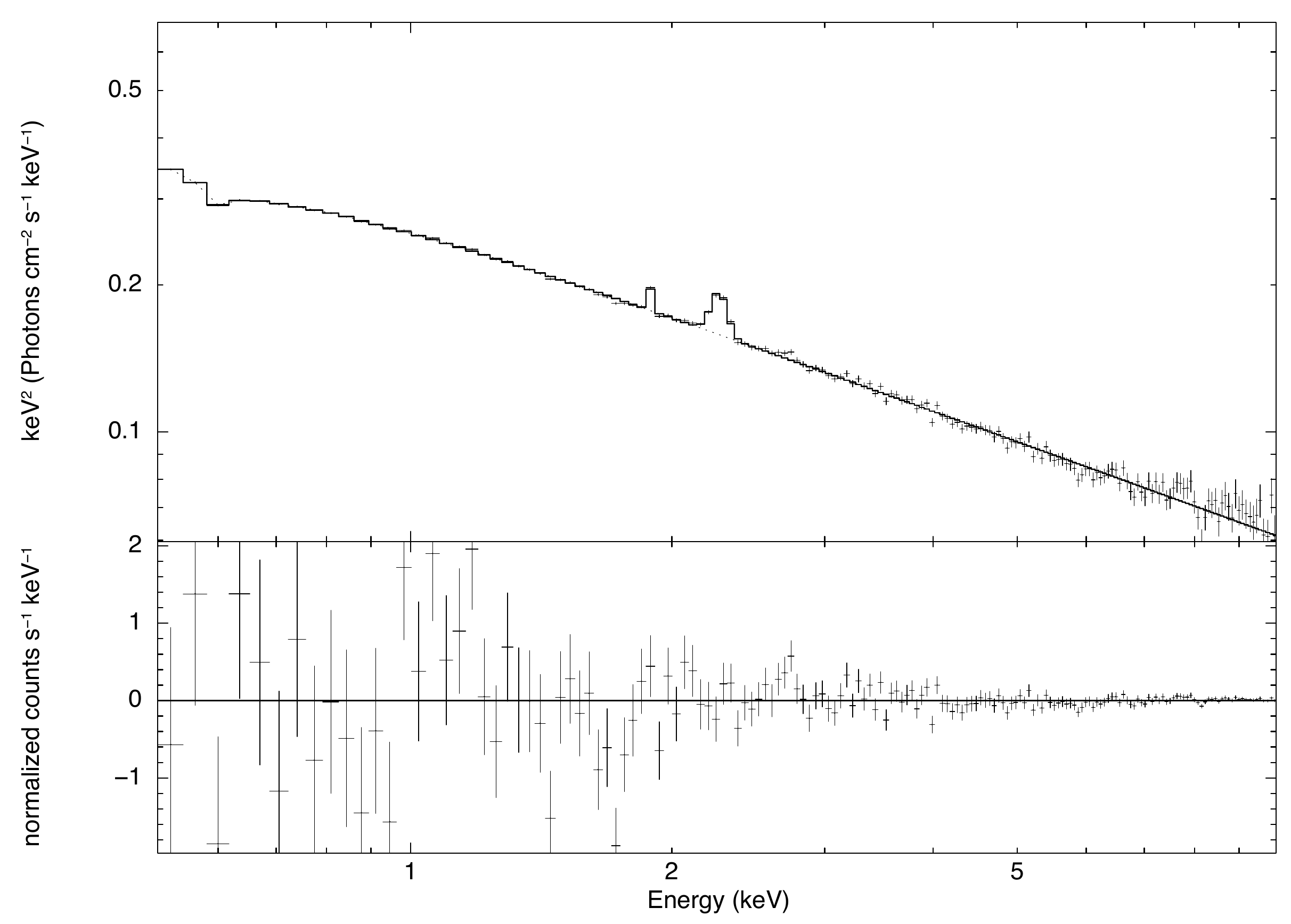}
                \label{0429pnspec}
}
        \quad
\subfloat[2014 May 1]{%
\includegraphics*[width=0.4\textwidth]{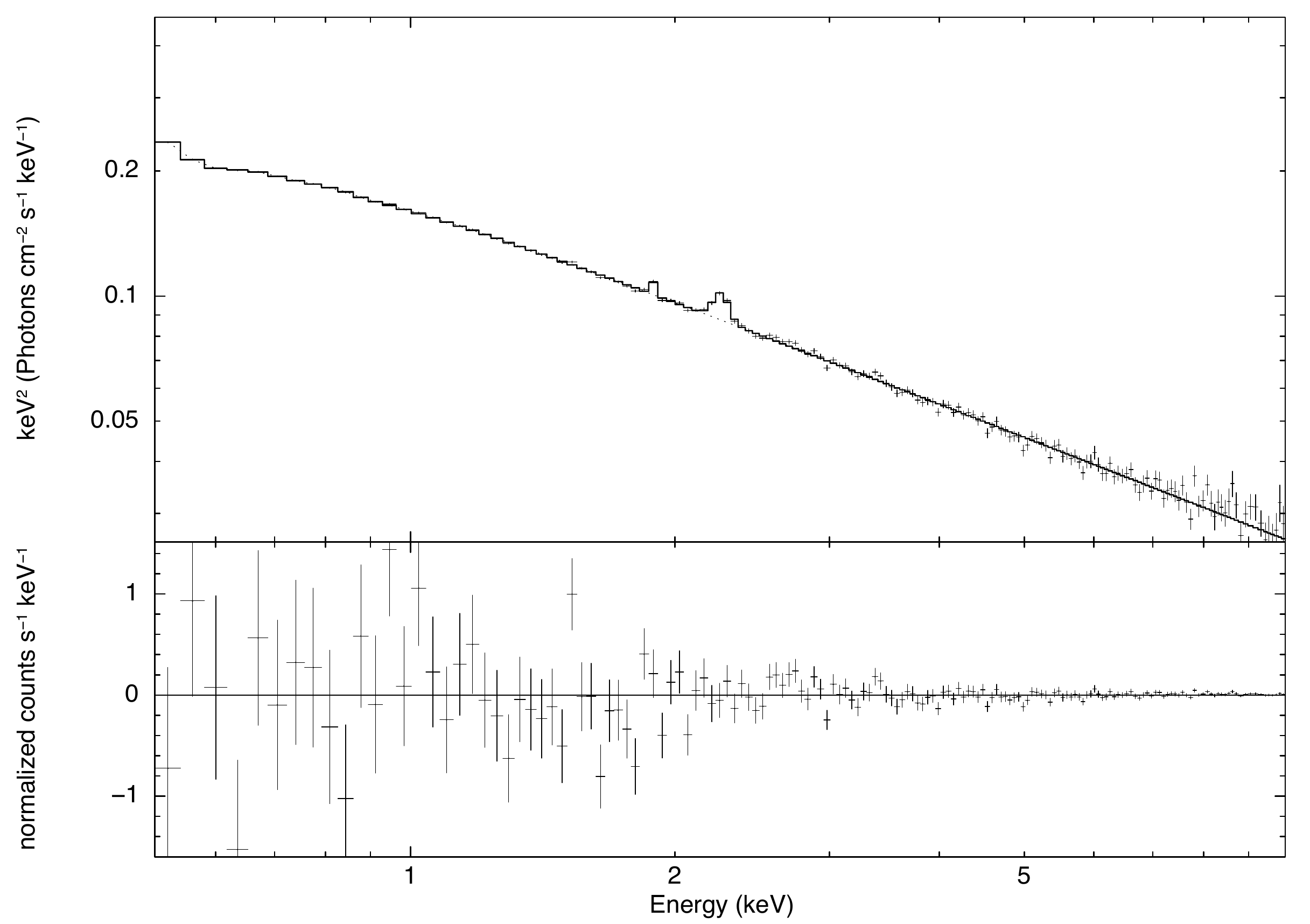}
                \label{0501pnspec}
}
        \quad
\subfloat[2014 May 3]{%
\includegraphics*[width=0.4\textwidth]{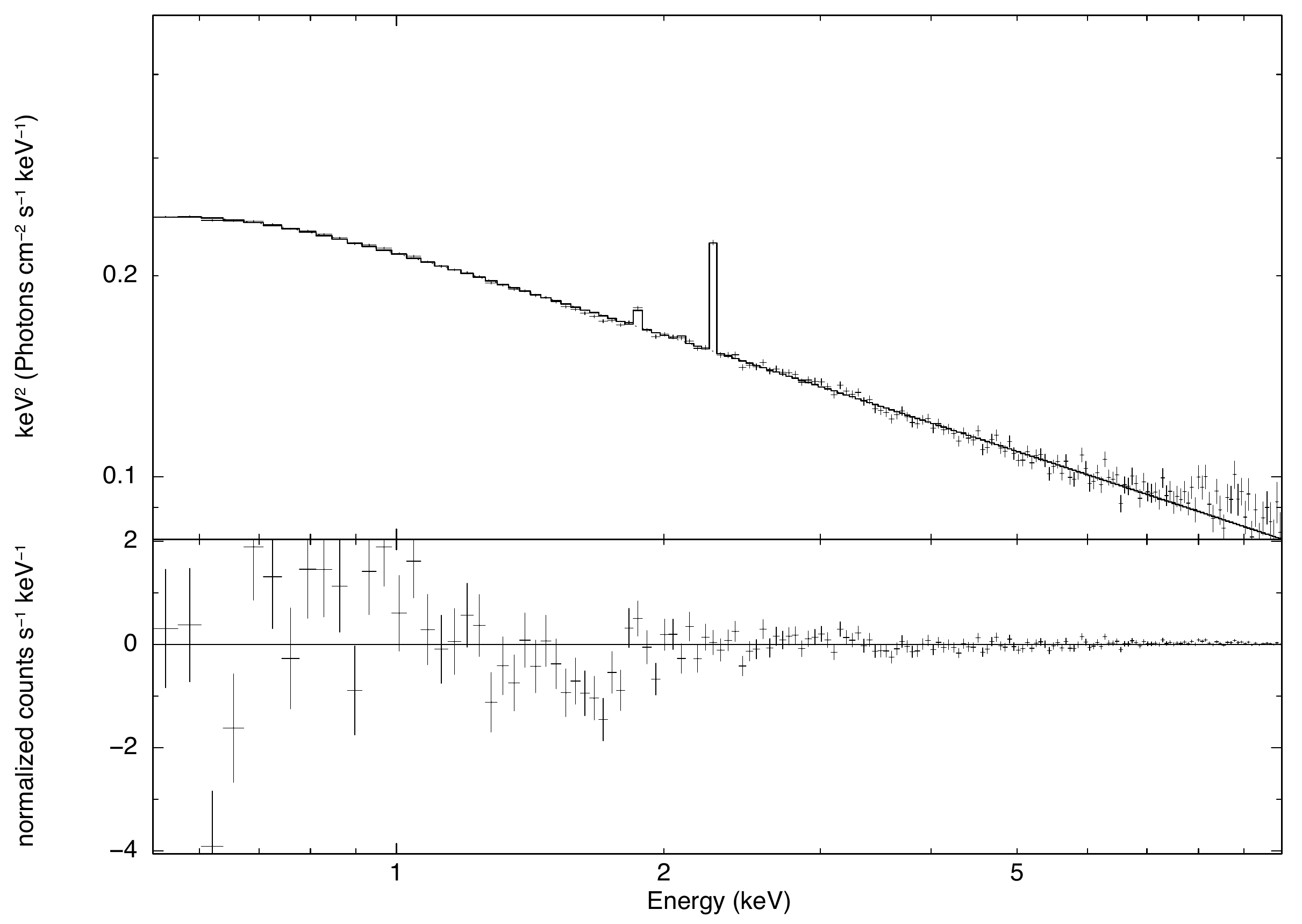}
                \label{0503pnspec}
}
\caption{X-ray spectra of Mrk 421 measured by {\it XMM-Newton}-EPN from the three simultaneous ToO observations in 2014. The spectra are fitted with a power law plus absorption model accounting for the source, and multiple instrumental features (see text and Table~\ref{XMMspec} for details). }
\label{pnspec}
\end{figure}
\begin{figure}[htp]
  \centering
    \includegraphics[width=1.0\textwidth]{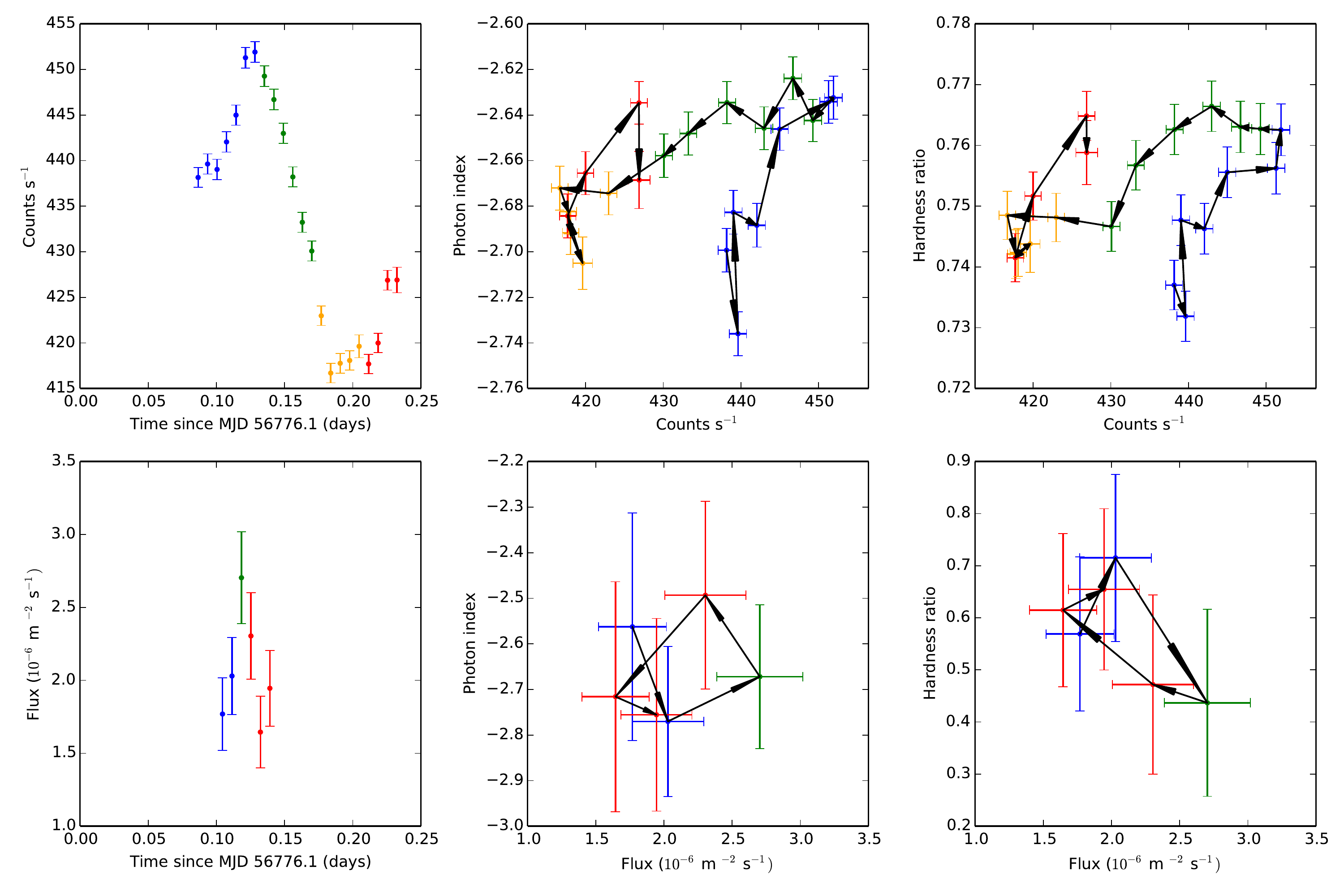}
    \caption{Spectral hysteresis of Mrk 421 on 2014 April 29. The top and bottom rows show results from X-ray and TeV observations, respectively. In each row, the left plot shows a light curve segment that contains a bump in flux, the middle plot shows the relationship between flux (counts) and best-fit photon index, and the right plot shows the relationship between flux (counts) and the hardness ratio. Each point of flux, HR, and index measurements is from a 10-min interval. The hardness ratio for X-ray is the ratio between the count rates in 1-10~keV and 0.5-1 keV; and for VHE between 560~GeV-30~TeV and 315-560~GeV. Black arrows and different colors are used to guide the eye as time progresses. }
    \label{0429hysteresis}
\end{figure}
\begin{figure}[htp]
  \centering
    \includegraphics[width=1.0\textwidth]{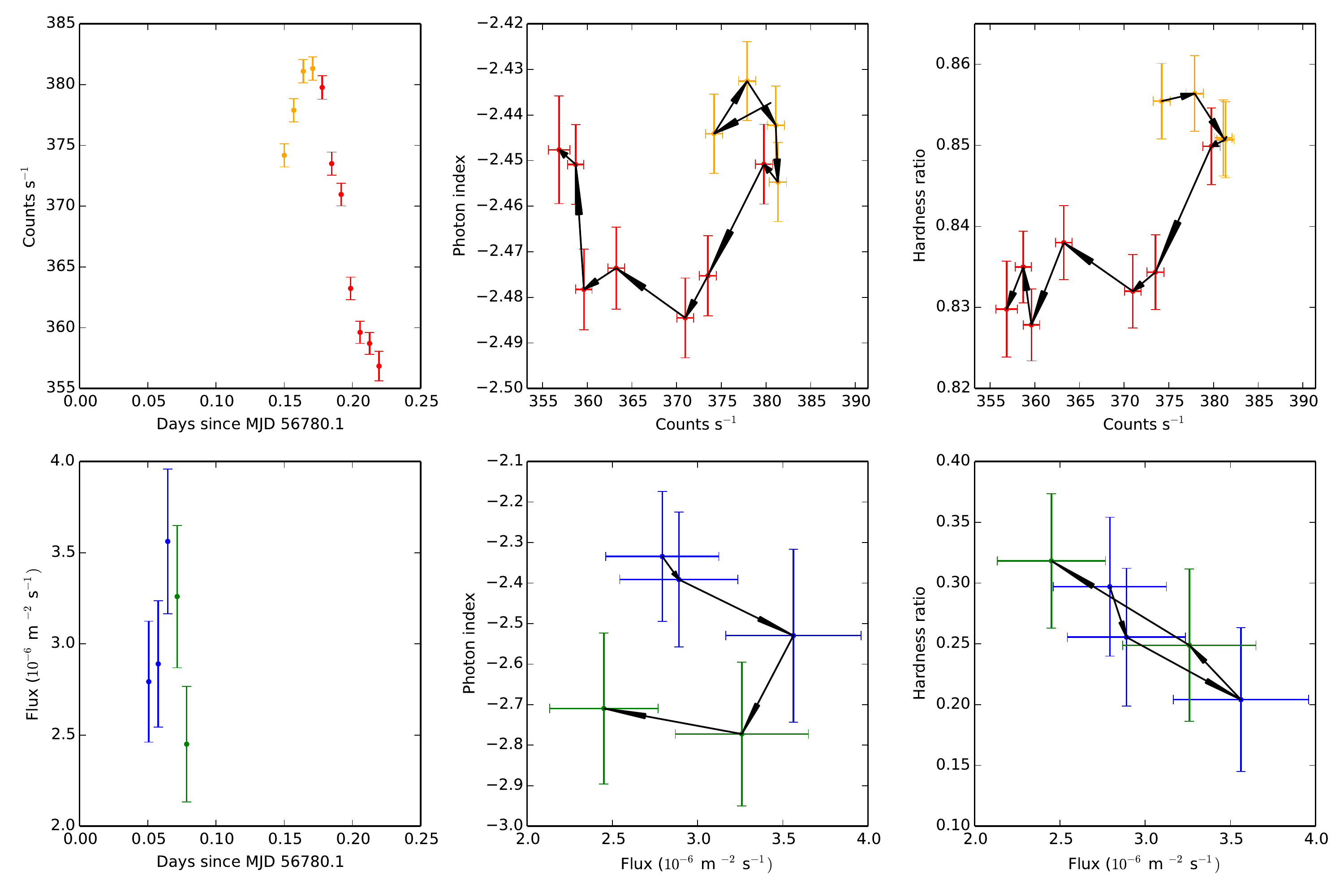}
    \caption{Spectral hysteresis of Mrk 421 on 2014 May 3. The top and bottom rows show results from X-ray and TeV observations, respectively. In each row, the left plot shows a light curve segment that contains a bump in flux, the middle plot shows the relationship between flux (counts) and best-fit photon index, and the right plot shows the relationship between flux (counts) and the hardness ratio. Each point of flux, HR, and index measurements is from a 10-min interval. The hardness ratio for X-ray is the ratio between the count rates in 1-10~keV and 0.5-1 keV; and for VHE between 560~GeV-30~TeV and 225-560~GeV. Black arrows and different colors are used to guide the eye as time progresses.  }
    \label{0503hysteresis}
\end{figure}
%

%
The nightly {\it XMM-Newton}-EPN X-ray spectra are fitted using a power-law model with absorption from neutral hydrogen, as well as instrumental features from oxygen, silicon, and gold, following the description in subsection~\ref{sec:XMM}. The results are shown in Figure~\ref{pnspec} and Table~\ref{XMMspec}. Note that the equivalent width $EW_i$ of each Gaussian component is shown instead of the standard deviation $\sigma_i$, because $\sigma_i$ is small compared to the energy-bin size of the X-ray spectrum and therefore not well-constrained by the fit. 

We then divide each of the {\it XMM-Newton} and VERITAS observations of Mrk~421 on the two nights with good atmospheric conditions into simultaneous 10-minute intervals, 
and perform spectral fitting for each interval, using an absorbed-power-law model for the X-ray data, and a power-law model for the gamma-ray data.  
We note that although the X-ray spectral fit is significantly improved by introducing the extra instrumental features, the spectral indices $\alpha$ remained unchanged within uncertainty. Therefore we are confident that the hardness ratios and the spectral indices derived for each 10-min interval are robust, and are not severely affected by the instrumental features. 


We identify several time intervals with a rise and a subsequent fall of flux in the X-ray light curves, and plot photon index and hardness ratio against flux (or count rate) for these bumps (see Figure~\ref{0429hysteresis} and Figure~\ref{0503hysteresis}). Black arrows indicate the order of time for each point. Measurements taken at different times are also color coded to guide the eye. 
A ``harder-when-brighter'' effect can be identified on some individual X-ray branches, (e.g. the blue and green points in the top right panel in Figure~\ref{0429hysteresis}). 
The observed ``soft lag" on May~3 indicates a harder spectrum when flux rises, and a softer spectrum when flux falls, corresponding to a clockwise loop (in orange color) in the top right panel of the spectral hysteresis plot in Figure~\ref{0503hysteresis}. Similarly, for the ``hard lag'' scenario on Apr~29, a counter-clockwise loop is predicted and observed, as shown in Figure~\ref{0429hysteresis}. 
It is interesting to note that the time lag and loop direction changes for flares separated by a few days, in spite of similar flux levels. 

The same analysis has been carried out for VHE data, and similar plots
are shown. The uncertainty in VHE flux, hardness ratio, index, and
hence hysteresis direction are large, and do not allow us to draw any
firm conclusions. 
We note that although the VHE spectrum of Mrk~421 is likely curved, a power-law model describes the data reasonably well for each 10-min interval (the average reduced $\chi^2$ value is $\sim$1.05). 

The hardness ratio pattern offers a more crude but
less model-dependent estimation of the same signature.  However, the
hardness-flux diagram of the VERITAS observations also has large
uncertainties.  At the flux level of roughly 1 to 2 C.~U., such
spectral hysteresis studies with the current generation of
ground-based gamma-ray instruments is difficult. This offers a
reference for the future in defining flux level trigger criteria for
target-of-opportunity observations, aiming for similar goals.


\subsection {Power spectral density}
The power spectral density (PSD) can be estimated from a periodogram, which is defined as the squared modulus of the Fourier transform of a time series 
\[
P(\nu)=\lvert F_N(\nu)\lvert^2=\left[\sum\limits_{i=1}^N f(t_i)cos(2\pi \nu t_i)\right]^2+\left[\sum\limits_{i=1}^N f(t_i)sin(2\pi \nu t_i)\right]^2. 
\]
We apply Leahy normalization to the periodogram to get the PSD, so that Poisson noise produces a constant power at an amplitude of 2. 

\begin{figure}[htp]
  \centering
\includegraphics[width=0.99\textwidth]{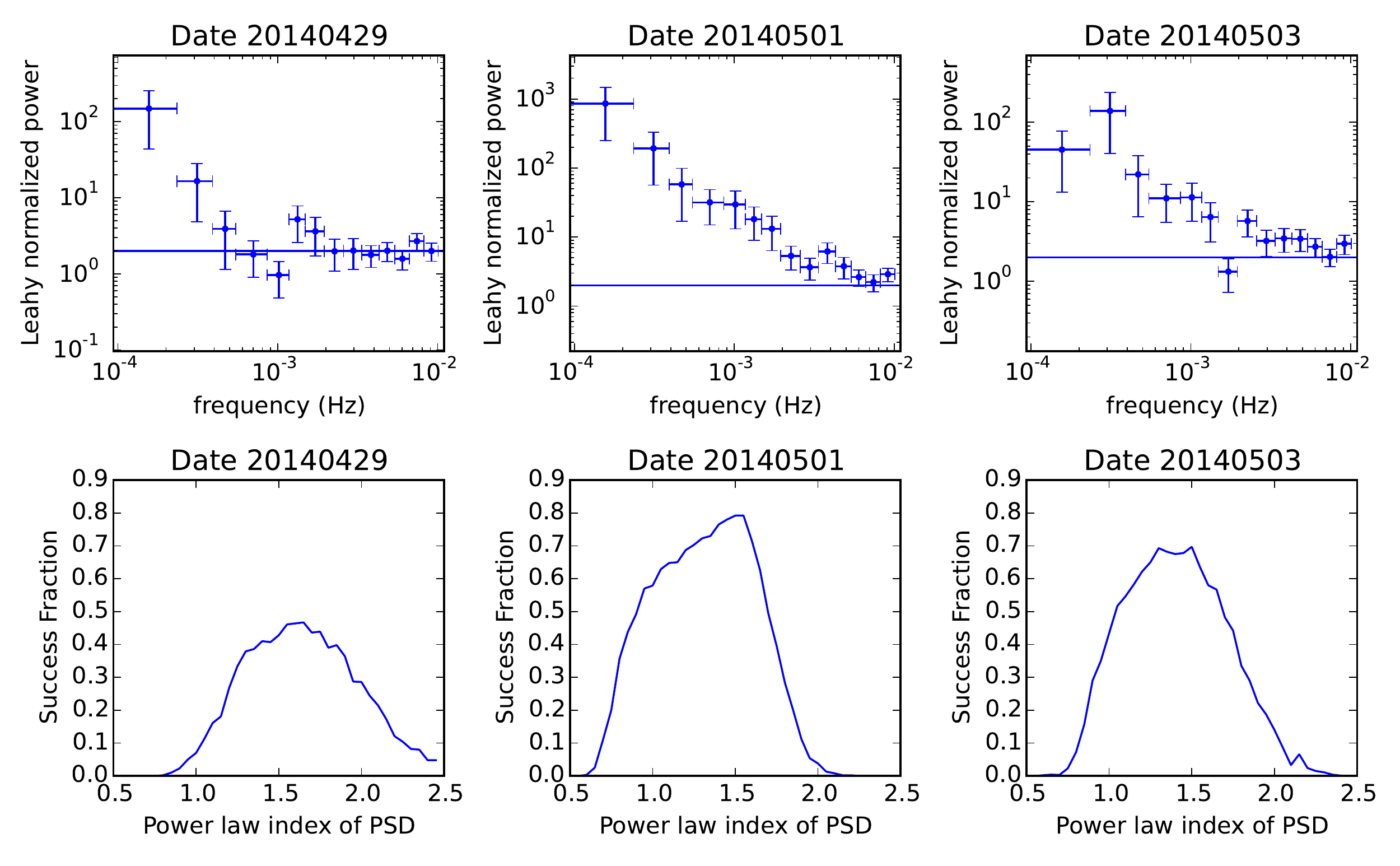}
\caption{ The PSD distributions (top row) and success fraction results (bottom row) of Mrk~421 calculated from the {\it XMM-Newton}-EPN observations in 2014. The power spectra are averaged over two segments of light curves each with 128 bins, and then rebinned geometrically in frequency space with a step factor 1.2. The PSDs are Leahy normalized, and the flat lines indicate the power of Poisson noise. The SuF results are calculated from comparisons between data and simulated light curves assuming an underlying PSD that follows a power-law distribution, the index of which goes from 0.5 to 2.5 in 0.05 steps. The highest frequencies ($>2\times10^{-3}$~Hz, $>7\times10^{-3}$~Hz, and $>3\times10^{-3}$~Hz on Apr 29, May 1, and May 3, respectively) are excluded when calculating the SuF to get rid of the bias caused by white noise. One thousand simulated light curves are generated at each index. }
\label{pnpsd}
\end{figure}

The top panels of Figure \ref{pnpsd} show PSDs calculated from the X-ray light curves of Mrk 421, measured by {\it XMM-Newton}-EPN on 2014 Apr 29, May 1, and May 3, respectively. 
The light curves are first binned by 50-s intervals, then divided into two equal-length segments each of 128 bins. A raw power spectrum is calculated for each segment, and averaged over both segments. Then, the power spectrum is rebinned geometrically with a step factor of 1.2 (i.e. a bin edge in frequency is the previous bin edge multiplied by a factor of 1.2). To estimate the error of the PSD in each bin, the standard deviation from all PSD points in the bin is used when it contains more than five points, otherwise the theoretical standard deviation of an exponential distribution is used. 
The PSDs cover a frequency range of $4\times 10^{-4}$ to $10^{-2}$~Hz. At higher frequency, the shape of the PSDs becomes flatter due to Poisson noise, which is shown as the flat line with a power of two under the Leahy normalization. 
However, we note that the PSD is well above the Poisson noise level up to $\sim$10$^{-3}$~Hz on all three days, which corresponds to timescales of under an hour. On May 1, the variability is still present, reaching $\sim2-3\times10^{-3}$~Hz, which is shorter than 10 minutes. 

We then simulate 1000 light curves at each of the indices ranging from 0.5 to 2.5 in steps of 0.1 following \citet{TK95}, and compare the data and the simulations to compute success fractions (SuF) as an estimation of the power-law index $\alpha$ following the method described by \citet{Uttley02}. %
The results are plotted in the bottom panels of Figure~\ref{pnpsd}. %
The SuF peaks at the PSD indices of $\sim$1.5 - 1.8 on Apr 29 with a peak value of $\sim$0.5, $\sim$1.3 - 1.6 on May~1 with a peak value of $\sim$0.8, and $\sim$1.2 - 1.6 on May~3 with a peak value of $\sim$0.7. These values represent the indices of the PSD better than the simple power-law fit the PSD, as SuFs are less biased by the spectral leakage which is present in both the data and the simulations. The range of PSD indices is consistent with previous studies of the same source at lower frequencies, 
e.g. a PSD index of 1.35 - 1.85 was found for frequency range $\sim 10^{-8} - 2 \times10^{-6}$ Hz  \citep{Isobe15}. 
However, a break at frequency of $\sim 9.5 \times 10^{-6}$~Hz was found in the X-ray PSD of Mrk 421 by \citet{Kataoka01}, and the PSD index above this frequency was determined to be 2.14. 
Surprisingly, we do not find evidence of such a steep PSD at $\gtrsim 4\times 10^{-4}$~Hz. 

\subsection{Broadband SEDs}
\label{sec:SED}
\begin{figure}[h]
  \centering
    \includegraphics[width=1.0\textwidth]{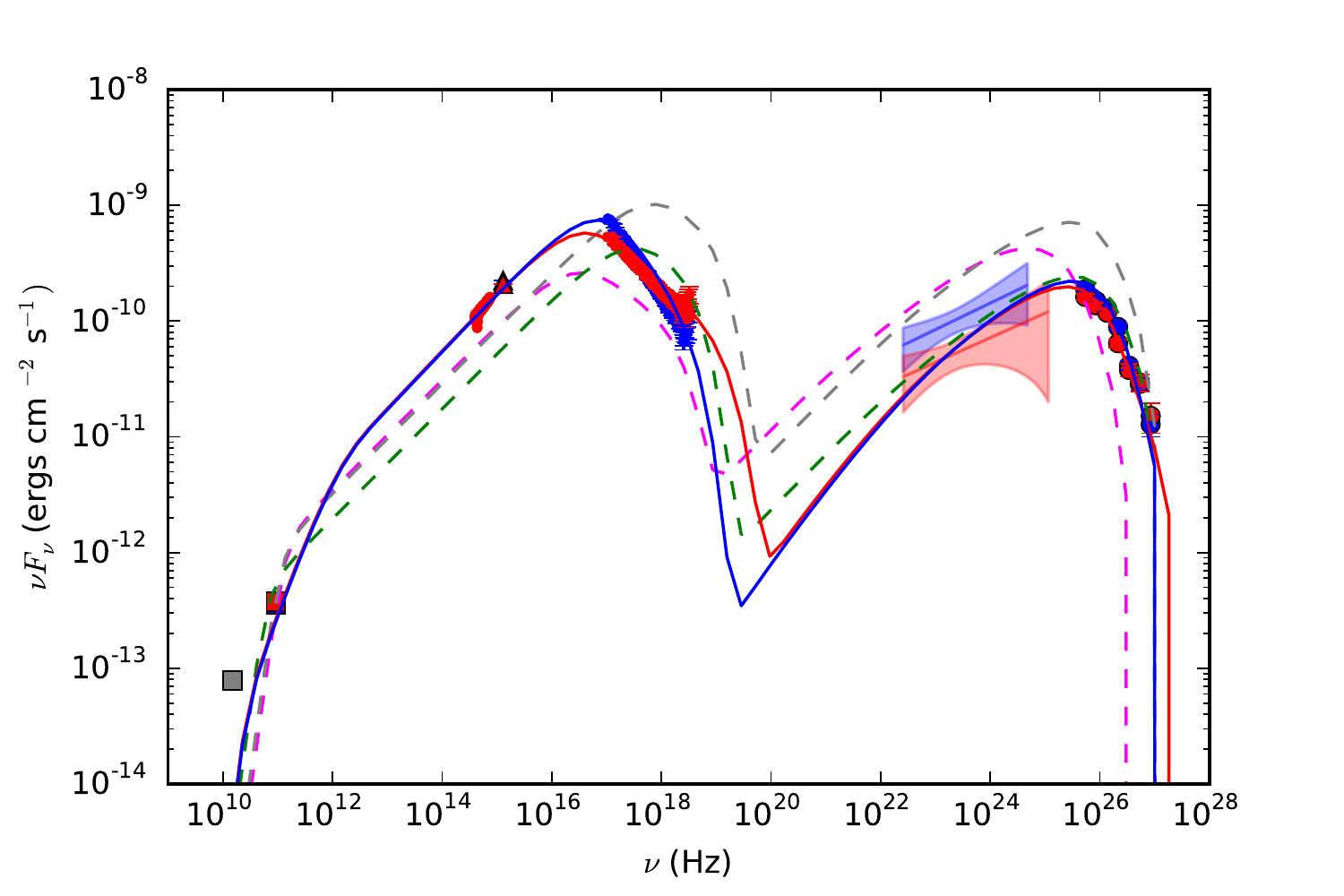}
    \caption{Broadband SED of Mrk 421 on 2014 April 29 (shown in blue) and May 3 (shown in red). From higher frequencies to lower ones, filled circles show VERITAS data, butterfly regions show {\it Fermi}-LAT data averaged over one day, round dots and triangles show {\it XMM-Newton}-EPN and OM data, respectively, small dots show Steward Observatory data (only on May 3, with host galaxy subtracted), and squares show CARMA and OVRO data. See text for details of the measurements and the SSC models shown. The results from previous observations are also shown for comparison: the gray, green, and magenta dashed lines correspond to models used for high, medium, and low flux as described in \citet{Blazejowski05}.  }
    \label{0429_0503SED}
\end{figure}

The SED of simultaneous VERITAS and {\it XMM-Newton} data, as well as contemporaneous MWL data, is shown in Figure~\ref{0429_0503SED}. 
Daily averaged high-energy (HE; $\sim$30~MeV - 100~GeV) gamma-ray spectra are constructed from {\it Fermi}-LAT data between 100 MeV and 300 GeV, and butterfly regions of 95\% confidence level are shown. Note that the uncertainty is large because of the scarcity of HE photons in the one-day window. Optical spectra from the Steward Observatory between 400 and 750 nm on May~3, radio data from CARMA at 95~GHz taken on both nights, and from OVRO at 15~GHz on other nights within the week are also shown. 

The X-ray and VHE spectra are located at the falling slopes of the lower- and higher-energy spectral peaks, respectively. 
The synchrotron peak is between the UV measurement at $\sim10^{15}$~Hz and the soft end of the X-ray spectrum at $\sim10^{17}$~Hz. Although the {\it Fermi}-LAT spectrum is not very constraining, the high-energy spectral peak appears to be just below $\sim$100~GeV, as suggested by the TeV spectrum. 
 
We use a static SSC model described in \citet{Krawczynski02} to study the observed SEDs. The set of parameters used to generate the solid blue and red curves as shown in Figure~\ref{0429_0503SED} are listed in Table~\ref{SSCmodel}. The static one-zone SSC model is roughly consistent with the data. %
The synchrotron peak frequency given by the model is $\nu_\text{syn}\sim4\times10^{16}$~Hz, while the inverse-Compton peak lies at $\nu_\text{SSC}\sim5\times10^{25}$~Hz. Using the peak frequencies and the observed spectral indices below and above the synchrotron peak, we can follow Equation 16 given by \citet{Tavecchio1998} assuming the system is in the Klein-Nishina (KN) regime and estimate the strength of the magnetic field to be 
\[
B \approx 2.7\times 10^{-7} \delta \frac{\nu_\text{syn}}{\nu_\text{SSC}^2} \left( \frac{mc^2}{h}\right)^2 \left\{ exp\left[  \frac{1}{\alpha_1-1}+\frac{1}{2(\alpha_2 - \alpha_1)} \right] \right\} ^2 \frac{1}{1+z} \approx 0.06 \;\text{G}, 
\]
where $\delta = [\Gamma (1-\beta cos \theta)]^{-1} \approx 20.3$ is the Doppler factor ($\Gamma$ and $\theta$ taken from the SED modeling results on May 3), $\alpha_1\sim0.5$ and $\alpha_2\sim1.5$ are the observed spectral indices below and above the synchrotron peak. 
We also calculate the Doppler factor limits for KN and Thomson regime to be $\sim$13 and $\sim$29, respectively, following Equation 13 and 17 in \citet{Tavecchio1998}. The Doppler factor value of 20.3 obtained from the SSC model is in between the two limits, therefore the gamma-ray emission is in the Thomson to KN transition regime. 

From Apr 29 to May 3, the change in SED might be described by an increase in the radius of the emitting region $R$, along with an increase in the maximum energy $E_{max}$, a slight decrease in break energy $E_{break}$ of the electron distribution, a harder electron spectrum (smaller $p_2$) after the break energy, and a slight increase in the Doppler factor (see Table~\ref{SSCmodel}). Note that a slight increase in magnetic field strength can have a similar effect as the increase in the Doppler factor. 
If these parameters indeed describe the evolution of the SED, it is consistent with the results of an expansion of the emitting region. The direct result of such an expansion is an increase in the dynamic timescale $t_{dyn}=R/c$. Moreover, this will lead to a higher maximum energy of the electrons $E_{max}$, since the maximum possible gyro-radius has increased. Also, the synchrotron cooling break, which occurs at the electron energy that satisfies $t_{syn}=t_{dyn}$, decreases since $t_{syn}\propto\gamma^{-1}$, where $\gamma$ is the Lorentz factor of the electron.

Under the simple SSC model, the hypothesis of an increase in the maximum energy $E_{max}$ would be consistent with a counter-clockwise spectral hysteresis pattern on Apr 29 and a clockwise pattern on May 3. The former is predicted when the system is observed near the maximum particle energy \citep[see e.g.][]{KRM98}, in which case the increasing acceleration timescale with higher particle energy results in the higher-energy photons lagging behind the lower-energy ones. Consequently, this leads to a softer spectrum when flux rises because lower-energy photons are emitted faster, and a counter-clockwise spectral hysteresis pattern. 
If the maximum electron energy $E_{max}$ is lower on Apr 29, the observed X-ray frequency (fixed at 0.5 - 10 keV) is closer to the maximum frequency, therefore the change in flux propagates from low to high frequencies and a counter-clockwise spectral hysteresis pattern should be observed. On the other hand, on May 3, as $E_{max}$ increases, the observed X-ray frequency covers a relatively lower portion of the entire particle spectrum, therefore the change in flux propagates from high to low energies and ``soft lag'' may be observed. 

\begin{landscape}
\begin{table}
   \caption{ Parameters used for the SSC model in Figure~\ref{0429_0503SED}. B05 refers to \citet{Blazejowski05}. 
   {\footnotesize $\Gamma$, $\theta$, and $R$ are the bulk Lorentz factor, the viewing angle, and the radius of the emitting region, respectively, $B$ is the strength of the magnetic field, $w_e$ is the energy density of the electrons, $\log E_{min}$, $\log E_{max}$, and $\log E_{break}$ are the logarithm of the minimum, maximum, and break electron energy, respectively, $p_1$ and $p_2$ are spectral index of the electrons below and above the break energy, respectively. }
   }
    \label{SSCmodel}
\begin{tabular}{cccccccccccc} \\ \hline \hline
Data set  & $\Gamma$ & $\theta$ & $B$ & $R$ & $w_e$ & $\log E_{min}[eV] $ & $\log E_{max}$[eV] & $\log E_{break}$[eV] & $p_1$ & $p_2$  \\
      &         & deg   &  $10^{-1}$ G & $10^{14} $ m      & ergs m$^{-3}$ &    &        &          &    & \\ \hline
  Apr 29 & 23.0 & 2.7 & 0.4 & 4.8 & 0.001 & 8.5 & 11.65 & 10.9 & 2.0 & 4.3 \\
  May 3 & 23.0 & 2.8 & 0.4 & 5.0 & 0.001 & 8.5 & 12.0 & 10.75 & 2.0 & 3.8 \\
  B05 low   & 19.48 & 5.01 & 4.05 & 0.7 & 0.13777 & 6.5 & 11.22 & 10.34 & 2.05 & 3.6  \\
  B05 med & 20.05 & 3.2 & 1.02 & 1.0 & 0.03192 & 6.5 & 11.55 & 10.98 & 2.05 & 3.4 \\
  B05 high  & 20.0 & 3.905 & 2.6 & 0.7 & 0.086 & 6.5 & 11.6 & 11.0 & 2.05 & 3.4 \\ \hline
\end{tabular}
\end{table}
\end{landscape}

\section{Discussion}
A detailed blazar variability study on sub-hour timescales using simultaneous and continuous VHE and X-ray observations has been presented in this work. Although it is challenging to carry out studies on such short timescales, they have the potential to constrain blazar models. Several different analyses in time and energy space are carried out, providing results that can crosscheck each other. Although the VHE flux level and dynamic range in this work are not sufficient to provide a conclusive picture of the emission mechanisms, the methods used are important for similar observations if the source is at higher flux levels and/or more variable. 

At the flux level of roughly $\sim$2~C.U., Mrk~421 shows $<20\%$ fractional variability in VHE gamma-ray flux on sub-hour to hour timescales, which seems low compared to previous studies on longer timescales at even lower fluxes. 
However, this is expected as the PSD of the variability follows a $1/f^{\alpha}$ style power law, leading to a lower variability power at higher frequencies (i.e. on shorter timescales) given that %
the value of $\alpha$ ranges from $\sim$1.2 to $\sim$1.8. For example, if a PSD of $1/f^{1.5}$ holds from $\sim1.65\times10^{-6}$~Hz (1 week) to 0.01~Hz (50 s), the fractional variability from timescales of 7 days down to 1 day would be $\sim$12 times higher than that from timescales probed in the {\it XMM-Newton} observations in this work. %
However, rare incidences of fast and strong variability in both X-ray and VHE have been observed before, e.g. variability with an amplitude of $\sim$15\% and on timescales of 20 minutes were reported by \citet{Blazejowski05}. Such fast flares could be the manifestation of $1/f^{\alpha}$ noise or individual local events caused by a different process. 
A steepening of the X-ray power spectral index to $\alpha \sim2.14$ at $\gtrsim 9.5 \times 10^{-6}$~Hz for Mrk~421 was reported by \citet{Kataoka01}. Such break features in the PSD of AGNs potentially carry information about the emission mechanism or characteristic timescales of the system. However, we do not find evidence of a much steeper PSD at $\gtrsim 4\times 10^{-4}$~Hz, compared with previous studies of the same source at lower frequencies. 

Several important timescales in blazars -- the cooling time $t_{cool}$, acceleration time $t_{acc}$, dynamic timescale $t_{dyn}$, and injection timescale $t_{inj}$ -- control many of the observable quantities, including the energy-dependent trends in their variability. %
For example, if the cooling timescale controls the flare timescale (the slow-cooling regime), a shorter decay timescale and greater fractional variability will be observed at higher energies, and a ``soft-lag'' and clockwise spectral hysteresis loop will be seen~\citep[e.g.,][]{KRM98}. 
``Soft-lag'' has been commonly observed in blazars in the X-ray band \citep[e.g.][]{Falcone04}, although ``hard-lag'' has also been reported \citep[e.g.,][]{Sato08}. Spectral hysteresis has also been observed from blazars, mostly from HBLs in the X-ray band \citep[e.g.,][]{Cui04}, as well as from a few flat-spectrum radio quasars in the HE gamma-ray band \citep[e.g.,][]{Nandikotkur07}). 


Within this work we attempt to measure spectral hysteresis patterns, as well as leads/lags between higher- and lower-energy bands, within both X-ray and VHE gamma-ray regimes, using the simultaneous ToO observations of Mrk 421 in these two bands in 2014. 
However, the lack of significant detection of time lags using two cross correlation methods suggests that, at the observed flux level (roughly between 1 and 2 C.~U.) and fractional variability amplitude ($\lesssim15\%$), the current ground-based gamma-ray instruments are still not sensitive enough for such studies. 
Future gamma-ray observations with current instruments at higher flux levels or with greater dynamic range, e.g. during a flare similar to the 10-C.~U. TeV gamma-ray flare from Mrk 421 that lasted for $\sim$1 hr \citep{Gaidos96}, or with more sensitive instruments, e.g. the Cherenkov Telescope Array, are needed to reach statistically significant conclusions about the sub-hour TeV variability of blazars. 

%
\acknowledgments
VERITAS is supported by grants from the U.S. Department of Energy Office of Science, the U.S. National Science Foundation and the Smithsonian Institution, and by NSERC in Canada. We acknowledge the excellent work of the technical support staff at the Fred Lawrence Whipple Observatory and at the collaborating institutions in the construction and operation of the instrument. 

The VERITAS Collaboration is grateful to Trevor Weekes for his seminal contributions and leadership in the field of VHE gamma-ray astrophysics, which made this study possible.

The MAGIC collaboration would like to thank
the Instituto de Astrof\'{\i}sica de Canarias
for the excellent working conditions
at the Observatorio del Roque de los Muchachos in La Palma.
The financial support of the German BMBF and MPG,
the Italian INFN and INAF,
the Swiss National Fund SNF,
the he ERDF under the Spanish MINECO
(FPA2015-69818-P, FPA2012-36668, FPA2015-68278-P,
FPA2015-69210-C6-2-R, FPA2015-69210-C6-4-R,
FPA2015-69210-C6-6-R, AYA2013-47447-C3-1-P,
AYA2015-71042-P, ESP2015-71662-C2-2-P, CSD2009-00064),
and the Japanese JSPS and MEXT
is gratefully acknowledged.
This work was also supported
by the Spanish Centro de Excelencia ``Severo Ochoa''
SEV-2012-0234 and SEV-2015-0548,
and Unidad de Excelencia ``Mar\'{\i}a de Maeztu'' MDM-2014-0369,
by grant 268740 of the Academy of Finland,
by the Croatian Science Foundation (HrZZ) Project 09/176
and the University of Rijeka Project 13.12.1.3.02,
by the DFG Collaborative Research Centers SFB823/C4 and SFB876/C3,
and by the Polish MNiSzW grant 745/N-HESS-MAGIC/2010/0.

This work used data from the {\it Fermi}-LAT archive and from the Steward Observatory spectropolarimetric monitoring project, which is supported by {\it Fermi} Guest Investigator grants NNX12AO93G and NNX15AU81G funded by NASA. 

This work made use of data supplied by the UK Swift Science Data Centre at the University of Leicester.

The OVRO 40-m monitoring program is supported in part by NASA grants NNX08AW31G and NNX11A043G, and NSF grants AST-0808050 and AST-1109911. 
M.\,B. acknowledges support from the International Fulbright Science and Technology Award, and NASA Headquarters under the NASA Earth and Space Science Fellowship Program, grant NNX14AQ07H.

\end{document}